\documentclass[10pt]{article}
\usepackage{amsmath}
\usepackage{amssymb}
\usepackage[cmtip,arrow]{xy}
\usepackage{pb-diagram,pb-xy}

\begin{document}

\title{\textbf{Double Lie algebras, semidirect product, and integrable
systems}\\
\smallskip\ }
\author{\textbf{S. Capriotti$^{\dag }$ \& H. Montani$^{\ddag }${\thanks{%
e-mail: \textit{\ hmontani@uaco.unpa.edu.ar }} } } \\
\dag\ \textit{Departamento de Matem\'{a}tica, Universidad Nacional del
Sur,\smallskip }\ \\
\ ~\textit{Av. Alem 1253, 8000} - \textit{Bah\'{\i}a Blanca, Buenos Aires,
Argentina. }\\
\\
\ddag\ CONICET \&\ \textit{Departamento de Ciencias Exactas y Naturales, }\\
~~\textit{Universidad Nacional de la Patagonia Austral.}\\
~~\textit{9011 - Caleta Olivia, Argentina}\\
}
\maketitle

\begin{abstract}
We study integrable systems on double Lie algebras in absence of
Ad-invariant bilinear form by passing to the semidirect product with the $%
\tau $-representation. We show that in this stage a natural Ad-invariant
bilinear form does exist, allowing for a straightforward application of the
AKS theory, and giving rise to Manin triple structure, thus bringing the
problem to the realm of Lie bialgebras and Poisson-Lie groups.
\end{abstract}


\newpage

\section{Introduction}

The deep relation between integrable systems and Lie algebras finds its
optimal realization when the involved Lie algebra is equipped with an
ad-invariant nondegenerate symmetric bilinear form. There, the coadjoint
orbit setting turns to be equivalent to the Lax pair formulation, the
Adler-Kostant-Symes theory of integrability \cite{Adler},\cite{Kostant},\cite%
{Symes} works perfectly, the Poisson-Lie group structures and Lie bialgebras
naturally arises, etc. Semisimplicity is an usual requirement warranting a
framework with this kind of bilinear form, however out of this framework it
becomes in a rather stringent condition. This is the case with semidirect
product Lie algebras.

Integrable systems can also be modelled on Lie groups, and their Hamiltonian
version is realized on their cotangent bundle. There, reduction of the
cotangent bundle of a Lie group by the action of some Lie subgroup brings
the problem to the realm of semidirect products \cite{Guillemin-Sternberg}, 
\cite{MWR} where, in spite of the lack of semisimplicity, an ad-invariant
form can be defined provided the original Lie algebra had one. Also, at the
Lie algebra level, in ref. \cite{Trofimov 1983} the complete integrability
of the Euler equations on a semidirect product of a semisimple Lie algebra
with its adjoint representation was proven. In ref. \cite{CapMon JPA}, the
AKS theory was applied to study integrable systems on this kind of Lie
groups.

However, the lack of an ad-invariant bilinear form is not an obstruction to
the application of AKS ideas. In fact, in ref. \cite{Ovando 1},\cite{Ovando
2} the AKS theory is adapted to a context equipped with a symmetric and
nondegenerate bilinear form, which also produces a decomposition of the Lie
algebra into two complementary orthogonal subspaces. This is performed by
using the \emph{B operation} introduced by Arnold in \cite{Arnold 1}, and
realizing that it amounts to be an action of the Lie algebra on itself,
which can be promoted to an action of the Lie group on its Lie algebra,
called the $\tau $-action. Thus, the restriction of the system to one of its
orthogonal components becomes integrable by factorization.

The main goal of this work is to study integrable systems on a semidirect
product of a Lie algebra with its adjoint representation, disregarding the
ad-invariance property of the bilinear form. So, the framework is that of a
double Lie algebra $\mathfrak{g}=\mathfrak{g}_{+}\oplus \mathfrak{g}_{-}$
equipped with a symmetric nondegenerate bilinear form, and the semidirect
product $\mathfrak{h}=\mathfrak{g}\ltimes _{\tau }\mathfrak{g}$ where the
left $\mathfrak{g}$ act on the other one by the $\tau $-action. The main
achievement is the introduction of an $\mathrm{ad}^{\mathfrak{h}}$-invariant
symmetric nondegenerate bilinear form which induces a decomposition $%
\mathfrak{h}=\mathfrak{h}_{+}\oplus \mathfrak{h}_{-}$, with $\mathfrak{h}%
_{+},\mathfrak{h}_{-}$ being Lie subalgebras and isotropic subspaces. In
this way, a natural Manin triple structure arises on the Lie algebra $%
\mathfrak{h}$, bringing the problem into the realm of the original AKS
theory and in that of Lie bialgebras and Poisson-Lie groups. In fact, we
show in which way integrable systems on $\mathfrak{h}_{\pm }$ arising from
the restriction of an almost trivial system on $\mathfrak{h}$ defined by an $%
\mathrm{ad}^{\mathfrak{h}}$-invariant Hamilton function, can be solved by
the factorization of an exponential curve in the associated connected
simply-connected Lie group $H$. Moreover, we built explicitly the
Poisson-Lie structures on the factors $H_{\pm }$ of the group $H$.

As the application of main interest, we think of that derived from Lie group
with no bi-invariant Riemannian metric. From the result by Milnor \cite%
{Milnor} asserting that a Riemannian metric is bi-invariant if and only if
the Lie group is a product of a compact semisimple and Abelian groups, one
finds a wide class of examples fitting in the above scheme among the
solvable and nilpotent Lie algebras. Many examples with dimension up to 6
are studied in ref. \cite{ghanam}, one of them is fully developed in the
present work as an example.

The work is ordered as follows: in Section II we fix the algebraic tools of
the problem by introducing the $\tau $-action, in Section III we present the
main results of this work, dealing with many issues in the semidirect
product framework. In Section IV we show how works the integrability by
factorization in the framework developed in the previous section. In Section
V, we present three examples without Ad-invariant bilinear forms on which we
apply the construction developed in the previous sections. Finally, in
Section VI we include some conclusions.

\section{Double Lie algebras and the $\protect\tau $-action}

Let us consider a \emph{double Lie group} $\left( G,G_{+},G_{-}\right) $ and
its associated \emph{double Lie algebra }$\left( \mathfrak{g},\mathfrak{g}%
_{+},\mathfrak{g}_{-}\right) $. These mean that $G_{+}$ and $G_{-}$ are Lie
subgroups of $G$ such that $G=G_{+}G_{-}$, and that $\mathfrak{g}_{+}$ and $%
\mathfrak{g}_{-}$ are Lie subalgebras of $\mathfrak{g}$ with $\mathfrak{g}=%
\mathfrak{g}_{+}\oplus \mathfrak{g}_{-}$. We also assume there is a
symmetric nondegenerate bilinear form $\left( \cdot ,\cdot \right) _{%
\mathfrak{g}}$ on $\mathfrak{g}$, which induces the direct sum
decomposition: 
\begin{equation*}
\mathfrak{g}=\mathfrak{g}_{+}^{\perp }\oplus \mathfrak{g}_{-}^{\perp }
\end{equation*}%
where $\mathfrak{g}_{\pm }^{\perp }$ are the annihilators subspaces of $%
\mathfrak{g}_{\pm }$, respectively, 
\begin{equation*}
\mathfrak{g}_{\pm }^{\perp }:=\left\{ Z\in \mathfrak{g}:\left( Z,X\right) _{%
\mathfrak{g}}=0\quad \forall X\in \mathfrak{g}_{\pm }\right\}
\end{equation*}

Since the bilinear form is not assumed to be $\mathrm{Ad}^{G}$-invariant,
the adjoint action is not a good symmetry in building integrable systems.
Following references $\cite{Ovando 1},\cite{Ovando 2}$, where AKS ideas are
adapted to a framework lacking of an ad-invariant bilinear form by using the
so called $\tau $\emph{-action}, we take this symmetry as the building block
of our construction. Let us briefly review the main result of these
references: let $\mathfrak{g}=\mathfrak{g}_{+}\oplus \mathfrak{g}_{-}$ as
above, then the adjoint action induces the $\tau $\emph{-action }defined as%
\begin{equation}
\begin{array}{ccc}
\mathrm{ad}^{\tau }:\mathfrak{g}\rightarrow \mathrm{End}\left( \mathfrak{g}%
\right) & / & \left( \mathrm{ad}_{X}^{\tau }Z,Y\right) _{\mathfrak{g}%
}:=-\left( Z,\left[ X,Y\right] \right) _{\mathfrak{g}}%
\end{array}
\label{ad-tao 0}
\end{equation}%
$\forall X,Y,Z\in \mathfrak{g}$. It can be promoted to an action of the
associated Lie group $G$ on $\mathfrak{g}$ through the exponential map,
thereby 
\begin{equation*}
\begin{array}{ccc}
\tau :G\rightarrow \mathrm{Aut}\left( \mathfrak{g}\right) & / & \left( \tau
\left( g\right) X,Y\right) _{\mathfrak{g}}:=\left( X,\mathrm{Ad}%
_{g^{-1}}^{G}Y\right) _{\mathfrak{g}}%
\end{array}%
\end{equation*}%
Often we also use the notation $\tau _{g}X=\tau \left( g\right) X$.

It is worth to observe that, since the bilinear form is nondegenerate, it
allows for the identification of the $\mathfrak{g}$ with its dual vector
space $\mathfrak{g}^{\ast }$ through the isomorphism 
\begin{equation}
\begin{array}{ccc}
\psi :\mathfrak{g}\longrightarrow \mathfrak{g}^{\ast } & / & \left\langle
\psi \left( X\right) ,Y\right\rangle :=\left( X,Y\right) _{\mathfrak{g}}%
\end{array}
\notag
\end{equation}%
which connects the $\tau $\emph{-action }with the \emph{coadjoint} one: 
\begin{equation*}
\tau \left( g\right) =\bar{\psi}\circ \mathrm{Ad}_{g^{-1}}^{G\ast }\circ \psi
\end{equation*}%
It reduces to the adjoint action, namely $\tau \left( g\right) =\mathrm{Ad}%
_{g}^{G}$, when the bilinear form is Ad-invariant. Also, $\psi $ provides
the isomorphisms $\psi :\mathfrak{g}_{\pm }^{\bot }\longrightarrow \mathfrak{%
g}_{\mp }^{\ast }$. Let $\Pi _{\mathfrak{g}_{\pm }^{\perp }}:\mathfrak{g}%
\rightarrow \mathfrak{g}_{\pm }^{\perp }$ be the projection map.

Observe that the annihilator subspaces $\mathfrak{g}_{\pm }^{\perp }$ can be
regarded as $\tau $\emph{-representation spaces }of $G_{\pm }$,
respectively. Moreover, the $\tau $\emph{-action }gives rise to crossed
actions, as explained in the following lemma.

\begin{description}
\item[Lemma:] \textit{The maps} $\tilde{\tau}:G_{\mp }\times \mathfrak{g}%
_{\pm }^{\perp }\rightarrow \mathfrak{g}_{\pm }^{\perp }$\textit{\ defined as%
}%
\begin{equation}
\tilde{\tau}\left( h_{\mp }\right) Z_{\pm }^{\perp }=\Pi _{\mathfrak{g}_{\pm
}^{\perp }}\left( \tau \left( h_{\mp }\right) Z_{\pm }^{\perp }\right)
\label{eq:NewAction}
\end{equation}%
\textit{are left actions, and the infinitesimal generator associated to
elements} $Y_{\mp }\in \mathfrak{g}_{\mp }$ \textit{is}%
\begin{equation*}
\left( Y_{\mp }\right) _{\mathfrak{g}_{\pm }^{\perp }}\left( Z_{\pm }^{\perp
}\right) =\Pi _{\mathfrak{g}_{\pm }^{\perp }}\left( \mathrm{ad}_{Y_{\mp
}}^{\tau }Z_{\pm }^{\perp }\right)
\end{equation*}
\end{description}

Thus, $\mathfrak{g}_{\pm }^{\perp }$ is a $G_{\mp }$-space through the $%
\mathrm{Ad}^{G\ast }$-action translated via the identification $\mathfrak{g}%
_{\pm }^{\perp }\simeq \mathfrak{g}_{\mp }^{\ast }$ induced by the inner
product. Further, this last identification allows to consider the
annihilators as Poisson manifolds, and the orbits of the $G_{\mp }$ on $%
\mathfrak{g}_{\pm }^{\perp }$ as symplectic manifolds. Under favorable
circumstances (i.e. when $f\in C^{\infty }\left( \mathfrak{g}\right) $ is $%
\mathrm{Ad}^{G}$-invariant) the dynamical system defined on these symplectic
spaces by the restriction of $f$ can be solved through the action of the
other group. As a bonus, the curve on this group whose action produces the
solution can be obtained by factorization of a simpler curve. Below we will
see the way in which it can be generalized to a context where there are no
Ad-invariant inner product.

\begin{description}
\item[Remark:] \textit{The assignment} $\mathrm{ad}^{\tau }:\mathfrak{g}%
\longrightarrow \mathrm{End}\left( \mathfrak{g}\right) $ \textit{is a Lie
algebra antihomomorphism}%
\begin{equation*}
\mathrm{ad}_{\left[ X,Y\right] }^{\tau }=-\left[ \mathrm{ad}_{X}^{\tau },%
\mathrm{ad}_{Y}^{\tau }\right]
\end{equation*}

\item[Remark:] $\cite{Ovando 1}$ \textit{The }$\tilde{\tau}$\textit{-orbits} 
$\mathcal{O}_{X}^{\tilde{\tau}}:=\left\{ \tilde{\tau}\left( h_{\mp }\right)
X\in \mathfrak{g}_{\pm }^{\perp }/h_{\mp }\in G_{\mp }\right\} $ \textit{%
whose tangent space are}%
\begin{equation*}
T_{\tau \left( g\right) X}\mathcal{O}_{X}^{\tau }=\left\{ \mathrm{ad}%
_{Y}^{\tau }\tau _{i}\left( g\right) X/Y\in \mathfrak{g}\right\}
\end{equation*}%
\textit{are symplectic manifold with the symplectic form}%
\begin{equation*}
\left\langle \omega ,\mathrm{ad}_{Y}^{\tau }\tau \left( g\right) X\otimes 
\mathrm{ad}_{Z}^{\tau }\tau \left( g\right) X\right\rangle _{\tau _{i}\left(
g\right) X}:=\left( \tau \left( g\right) X,\left[ Y,Z\right] \right) _{%
\mathfrak{g}}
\end{equation*}
\end{description}

\section{Semidirect product with the $\protect\tau $-action representation}

The lack of an $\mathrm{Ad}$-invariant bilinear form on $\mathfrak{g}=%
\mathfrak{g}_{+}\oplus \mathfrak{g}_{-}$ is not an obstacle to have AKS
integrable systems in the \emph{double Lie algebra }$\left( \mathfrak{g},%
\mathfrak{g}_{+},\mathfrak{g}_{-}\right) $ $\cite{Ovando 1}$. However, it
prevents the existence of richer structures as the \emph{Manin triple} one,
which brings it to the realm of \emph{Lie bialgebras} and the associated Lie
groups acquire a natural compatible Poisson structure turning them into 
\emph{Poisson-Lie groups}.

In this section we introduce the semidirect sum Lie algebra $\mathfrak{h}=%
\mathfrak{g}\ltimes _{\tau }\mathfrak{g}$ as the central object of our
construction.

\subsection{Lie algebra structure on $\mathfrak{h}=\mathfrak{g}\oplus 
\mathfrak{g}$ and ad-invariant bilinear form}

Let us consider the semidirect sum Lie algebra $\mathfrak{h}=\mathfrak{g}%
\ltimes _{\tau }\mathfrak{g}$ where the first component acts on the second
one through the $\mathrm{ad}^{\tau }$-action $\left( \ref{ad-tao 0}\right) $%
, giving rise to the Lie bracket on $\mathfrak{h}$ 
\begin{equation}
\left[ \left( X,Y\right) ,\left( X^{\prime },Y^{\prime }\right) \right]
:=\left( \left[ X,X^{\prime }\right] ,\mathrm{ad}_{X}^{\tau }Y^{\prime }-%
\mathrm{ad}_{X^{\prime }}^{\tau }Y\right)  \label{Lie alg g+g 1}
\end{equation}%
Also, we equip $\mathfrak{h}$ with the symmetric nondegenerate bilinear form 
$\left( -,-\right) _{\mathfrak{h}}:\mathfrak{h}\times \mathfrak{h}%
\rightarrow \mathbb{R}$%
\begin{equation}
\left( \left( X,Y\right) ,\left( X^{\prime },Y^{\prime }\right) \right) _{%
\mathfrak{h}}:=\left( X,Y^{\prime }\right) _{\mathfrak{g}}+\left(
Y,X^{\prime }\right) _{\mathfrak{g}}  \label{Lie alg g+g 2}
\end{equation}%
It is easy to prove the next result.

\begin{description}
\item[Proposition:] \textit{The bilinear form }$\left( \cdot ,\cdot \right)
_{\mathfrak{h}}\ $\textit{on the semidirect sum of Lie algebras }$\mathfrak{h%
}$ \textit{is }$\mathit{\mathrm{ad}}^{\mathfrak{h}}$\textit{-invariant}%
\begin{equation*}
\left( \left[ \left( X^{\prime \prime },Y^{\prime \prime }\right) ,\left(
X,Y\right) \right] ,\left( X^{\prime },Y^{\prime }\right) \right) _{%
\mathfrak{h}}+\left( \left( X,Y\right) ,\left[ \left( X^{\prime \prime
},Y^{\prime \prime }\right) ,\left( X^{\prime },Y^{\prime }\right) \right]
\right) _{\mathfrak{h}}=0
\end{equation*}
\end{description}

Let us denote by $H=G\ltimes _{\tau }\mathfrak{g}$ the Lie group associated
with the above Lie algebra structure. Indeed, there are two possible
semidirect product Lie group structures related to the left or right
character of the action of $G$ on $\mathfrak{g}$. We adopt the right action
structure for the Lie group structure in $H=G\times \mathfrak{g}$%
\begin{equation}
\left( g,X\right) \mathbf{\cdot }\left( k,Y\right) :=\left( gk,\tau
_{k^{-1}}X+Y\right)  \label{semidirect prod}
\end{equation}%
The exponential map in this case is 
\begin{equation}
\mathrm{Exp}^{\cdot }\left( t\left( X,Y\right) \right) =\left(
e^{tX},-\left( \sum_{n=1}^{\infty }\frac{\left( -1\right) ^{n}}{n!}%
t^{n}\left( \mathrm{ad}_{X}^{\tau }\right) ^{n-1}\right) Y\right)
\label{exp semid transad}
\end{equation}%
and the adjoint action of $H$ on $\mathfrak{h}$ is%
\begin{equation*}
\mathrm{Ad}_{\left( g,Z\right) }^{H}\left( X,Y\right) =\left( \mathrm{Ad}%
_{g}^{G}X\,,\tau _{g}\left( Y-\mathrm{ad}_{X}^{\tau }Z\right) \right)
\end{equation*}%
while the adjoint action of $\mathfrak{h}$ on itself retrieves de Lie
bracket structure $\left( \ref{Lie alg g+g 1}\right) $.

\subsection{Manin triple and factorization}

The direct sum decompositions of $\mathfrak{g}$ as $\mathfrak{g}=\mathfrak{g}%
_{+}\oplus \mathfrak{g}_{-}$ and $\mathfrak{g}=\mathfrak{g}_{+}^{\perp
}\oplus \mathfrak{g}_{-}^{\perp }$ allows to decompose the Lie algebra $%
\mathfrak{h}$ as a direct sum $\mathfrak{h}=\mathfrak{h}_{+}\oplus \mathfrak{%
h}_{-}$, where $\mathfrak{h}_{+},\mathfrak{h}_{-}$ are Lie subalgebras of $%
\mathfrak{h}$. In fact, by observing that the subspaces $\mathfrak{g}_{\pm
}^{\bot }$ are $\tau $\emph{-representation spaces }of $G_{\pm }$, we define
the semidirect products 
\begin{equation*}
H_{\pm }=G_{\pm }\ltimes _{\tau }\mathfrak{g}_{\pm }^{\bot }
\end{equation*}%
with 
\begin{equation*}
\left( g_{\pm },X_{\pm }^{\bot }\right) \cdot \left( k_{\pm },Y_{\pm }^{\bot
}\right) :=\left( g_{\pm }k_{\pm },\tau _{k_{\pm }^{-1}}X_{\pm }^{\bot
}+Y_{\pm }^{\bot }\right) ~,
\end{equation*}%
and the semidirect sum Lie algebras 
\begin{equation*}
\mathfrak{h}_{\pm }=\mathfrak{g}_{\pm }\ltimes _{\tau }\mathfrak{g}_{\pm
}^{\bot }
\end{equation*}%
with the Lie bracket 
\begin{equation}
\left[ \left( X_{\pm },Y_{\pm }^{\bot }\right) ,\left( X_{\pm }^{\prime
},Y_{\pm }^{\prime \bot }\right) \right] =\left( \left[ X_{\pm },X_{\pm
}^{\prime }\right] ,\mathrm{ad}_{X_{\pm }}^{\tau }Y_{\pm }^{\prime \bot }-%
\mathrm{ad}_{X_{\pm }^{\prime }}^{\tau }Y_{\pm }^{\bot }\right) \in 
\mathfrak{h}_{_{\pm }}~.  \label{Lie subalg}
\end{equation}%
Then, we have the factorization 
\begin{equation*}
H=H_{+}H_{-}~,
\end{equation*}%
and the decomposition in direct sum of Lie subalgebras 
\begin{equation*}
\mathfrak{h}=\mathfrak{h}_{+}\oplus \mathfrak{h}_{-}~.
\end{equation*}

In addition, the restriction of the bilinear form $\left( \ref{Lie alg g+g 2}%
\right) $ to the subspaces $\mathfrak{h}_{+},\mathfrak{h}_{-}$ vanish 
\begin{equation*}
\left( \left( X_{\pm },Y_{\pm }^{\bot }\right) ,\left( X_{\pm }^{\prime
},Y_{\pm }^{\prime \bot }\right) \right) _{\mathfrak{h}}=\left( X_{\pm
},Y_{\pm }^{\prime \bot }\right) _{\mathfrak{g}}+\left( Y_{\pm }^{\bot
},X_{\pm }^{\prime }\right) _{\mathfrak{g}}=0
\end{equation*}%
meaning that the $\mathfrak{h}_{+},\mathfrak{h}_{-}$ are isotropic subspaces
of $\mathfrak{h}$. These results are summarized in the following proposition.

\begin{description}
\item[Proposition:] \textit{The Lie algebras }$\mathfrak{h}$, $\mathfrak{h}%
_{_{+}}$, $\mathfrak{h}_{_{-}}$ \textit{endowed with the bilinear form }$%
\left( \ref{Lie alg g+g 2}\right) $ \textit{compose a Manin triple }$\left( 
\mathfrak{h},\mathfrak{h}_{_{+}},\mathfrak{h}_{_{-}}\right) $.
\end{description}

Hence, the Lie subalgebras $\mathfrak{h}_{_{\pm }}=\mathfrak{g}_{\pm
}\ltimes _{\tau }\mathfrak{g}_{\pm }^{\bot }$ are a Lie bialgebras, and the
factors $H_{_{\pm }}=G_{\pm }\ltimes _{\tau }\mathfrak{g}_{\pm }^{\bot }$
are Poisson-Lie groups.

The factorization $H=H_{+}H_{-}$ means that each $\left( g,X\right) \in H$
can be written as 
\begin{equation}
\left( g,X\right) =\left( g_{+},\tilde{\tau}_{g_{-}}\left( \Pi _{\mathfrak{g}%
_{+}^{\bot }}X\right) \right) \cdot \left( g_{-},\Pi _{\mathfrak{g}%
_{-}^{\bot }}\left( Id-\tau _{g_{-}^{-1}}\tilde{\tau}_{g_{-}}\Pi _{\mathfrak{%
g}_{+}^{\bot }}\right) X\right)  \label{semidirect fact 2}
\end{equation}%
with $\tilde{\tau}:G_{\mp }\times \mathfrak{g}_{\pm }^{\perp }\rightarrow 
\mathfrak{g}_{\pm }^{\perp }$\textit{\ }defined in $\left( \ref{eq:NewAction}%
\right) $.

Let us name $\gamma :\mathfrak{h}\longrightarrow \mathfrak{h}^{\ast }$ the
linear bijection induced by the bilinear form $\left( ,\right) _{\mathfrak{h}%
}$, then it also produces the linear bijections $\gamma \left( \mathfrak{h}%
_{_{\pm }}\right) =\mathfrak{h}_{_{\mp }}^{\ast }$.

\subsection{Dressing action}

Following ref. $\cite{Lu-We}$, we write the Lie bracket in the double Lie
algebra $\left( \mathfrak{h},\mathfrak{h}_{+},\mathfrak{h}_{-}\right) $ as 
\begin{equation*}
\left[ \left( X_{-},Y_{-}^{\bot }\right) ,\left( X_{+},Y_{+}^{\bot }\right) %
\right] =\left( X_{+},Y_{+}^{\bot }\right) ^{\left( X_{-},Y_{-}^{\bot
}\right) }+\left( X_{-},Y_{-}^{\bot }\right) ^{\left( X_{+},Y_{+}^{\bot
}\right) }
\end{equation*}%
with $\left( X_{+},Y_{+}^{\bot }\right) ^{\left( X_{-},Y_{-}^{\bot }\right)
}\in \mathfrak{h}_{+}$ and $\left( X_{+},Y_{+}^{\bot }\right) ^{\left(
X_{-},Y_{-}^{\bot }\right) }\in \mathfrak{h}_{-}$. Therefore, from the Lie
algebra structure $\left( \ref{Lie alg g+g 1}\right) $ we get%
\begin{equation}
\left\{ 
\begin{array}{l}
\left( X_{-},Y_{-}^{\bot }\right) ^{\left( X_{+},Y_{+}^{\bot }\right)
}=\left( X_{-}^{X_{+}},\Pi _{\mathfrak{g}_{-}^{\bot }}\mathrm{ad}%
_{X_{-}}^{\tau }Y_{+}^{\bot }-\Pi _{\mathfrak{g}_{-}^{\bot }}\mathrm{ad}%
_{X_{+}}^{\tau }Y_{-}^{\bot }\right) \\ 
\\ 
\left( X_{+},Y_{+}^{\bot }\right) ^{\left( X_{-},Y_{-}^{\bot }\right)
}=\left( X_{+}^{X_{-}},\Pi _{\mathfrak{g}_{+}^{\bot }}\mathrm{ad}%
_{X_{-}}^{\tau }Y_{+}^{\bot }-\Pi _{\mathfrak{g}_{+}^{\bot }}\mathrm{ad}%
_{X_{+}}^{\tau }Y_{-}^{\bot }\right)%
\end{array}%
\right.  \label{Lie alg dress action}
\end{equation}

Let us consider the factorization for $\left( g,X\right) \in H$ given in $%
\left( \ref{semidirect fact 2}\right) $, and any couple $\left(
g_{+},X_{+}^{\bot }\right) \in H_{+}$, $\left( g_{-},X_{-}^{\bot }\right)
\in H_{-}$. Since the product $\left( g_{-},X_{-}^{\bot }\right) \cdot
\left( g_{+},X_{+}^{\bot }\right) $ is also in $H$, it can be decomposed as
above. Then, we write 
\begin{equation*}
\left( g_{-},X_{-}^{\bot }\right) \cdot \left( g_{+},X_{+}^{\bot }\right)
=\left( g_{+},X_{+}^{\bot }\right) ^{\left( g_{-},X_{-}^{\bot }\right)
}\cdot \left( g_{-},X_{-}^{\bot }\right) ^{\left( g_{+},X_{+}^{\bot }\right)
}
\end{equation*}%
for $\left( g_{+},X_{+}\right) ^{\left( g_{-},X_{-}\right) }\in H_{+}$ and $%
\left( g_{-},X_{-}\right) ^{\left( g_{+},X_{+}\right) }\in H_{-}$ given by 
\begin{equation*}
\left\{ 
\begin{array}{l}
\left( g_{+},X_{+}^{\bot }\right) ^{\left( g_{-},X_{-}^{\bot }\right)
}=\left( g_{+}^{g_{-}},\tilde{\tau}_{g_{-}^{g_{+}}}\left( \Pi _{\mathfrak{g}%
_{+}^{\bot }}\tau _{g_{+}^{-1}}X_{-}^{\bot }+X_{+}^{\bot }\right) \right) \\ 
\\ 
\left( g_{-},X_{-}^{\bot }\right) ^{\left( g_{+},X_{+}^{\bot }\right) } \\ 
\qquad =\left( g_{-}^{g_{+}},\Pi _{\mathfrak{g}_{-}^{\bot }}\tau
_{g_{+}^{-1}}X_{-}^{\bot }-\Pi _{\mathfrak{g}_{-}^{\bot }}\tau _{\left(
g_{-}^{g_{+}}\right) ^{-1}}\tilde{\tau}_{g_{-}^{g_{+}}}\left( X_{+}^{\bot
}+\Pi _{\mathfrak{g}_{+}^{\bot }}\tau _{g_{+}^{-1}}X_{-}^{\bot }\right)
\right)%
\end{array}%
\right.
\end{equation*}%
The assignment $H_{-}\times H_{+}\longrightarrow H_{+}$ and $H_{+}\times
H_{-}\longrightarrow H_{-}$ defined by the above relations are indeed \emph{%
actions}, the so-called \emph{dressing actions }(see \cite{STS} and \cite%
{Lu-We}). In fact, $H_{-}\times H_{+}\longrightarrow H_{+}$ such that $%
\left( \left( g_{-},X_{-}^{\bot }\right) ,\left( g_{+},X_{+}^{\bot }\right)
\right) \longmapsto \left( g_{+},X_{+}^{\bot }\right) ^{\left(
g_{-},X_{-}^{\bot }\right) }$ is a \emph{right action} of $H_{-}$ on $H_{+}$
and, reciprocally, $\left( \left( g_{+},X_{+}^{\bot }\right) ,\left(
g_{-},X_{-}^{\bot }\right) \right) \longmapsto \left( g_{-},X_{-}^{\bot
}\right) ^{\left( g_{+},X_{+}^{\bot }\right) }$ is a \emph{left action} of $%
H_{+}$ on $H_{-}$.

The infinitesimal generator of the dressing actions can be derived be
considering the action of exponential elements $\left( g_{-},X_{-}^{\bot
}\right) =\mathrm{Exp}^{\cdot }\left( t\left( X_{-},Y_{-}^{\bot }\right)
\right) $ and $\left( g_{+},X_{+}^{\bot }\right) =\mathrm{Exp}^{\cdot
}\left( t\left( X_{+},Y_{+}^{\bot }\right) \right) $ to get%
\begin{equation}
\left\{ 
\begin{array}{l}
\left( g_{+},X_{+}^{\bot }\right) ^{\left( X_{-},Y_{-}^{\bot }\right)
}=\left( g_{+}^{X_{-}},\Pi _{\mathfrak{g}_{+}^{\bot }}\mathrm{ad}%
_{X_{-}^{g_{+}}}^{\tau }X_{+}^{\bot }+\Pi _{\mathfrak{g}_{+}^{\bot }}\tau
_{g_{+}^{-1}}Y_{-}^{\bot }\right) \\ 
\\ 
\left( g_{-},X_{-}^{\bot }\right) ^{\left( X_{+},Y_{+}^{\bot }\right) } \\ 
\qquad =\left( g_{-}^{X_{+}},\Pi _{\mathfrak{g}_{-}^{\bot }}\tau
_{g_{-}^{-1}}\tilde{\tau}_{g_{-}}\left( \Pi _{\mathfrak{g}_{+}^{\bot }}%
\mathrm{ad}_{X_{+}}^{\tau }X_{-}^{\bot }-Y_{+}^{\bot }\right) -\Pi _{%
\mathfrak{g}_{-}^{\bot }}\mathrm{ad}_{X_{+}}^{\tau }X_{-}^{\bot }\right)%
\end{array}%
\right.  \label{inf gen dres transad}
\end{equation}%
and making the derivative at $t=0$, we recover the result obtained in $%
\left( \ref{Lie alg dress action}\right) $.

Also one may show that 
\begin{equation*}
\left\{ 
\begin{array}{l}
\left( X_{+},X_{+}^{\bot }\right) ^{\left( g_{-},X_{-}^{\bot }\right)
}=\left( X_{+}^{g_{-}},\tilde{\tau}_{g_{-}}X_{+}^{\bot }-\tilde{\tau}%
_{g_{-}}\Pi _{\mathfrak{g}_{+}^{\bot }}\mathrm{ad}_{X_{+}}^{\tau
}X_{-}^{\bot }\right) \\ 
\\ 
\left( X_{-},X_{-}^{\bot }\right) ^{\left( g_{+},X_{+}^{\bot }\right)
}=\left( X_{-}^{g_{+}},\tilde{\tau}_{g_{+}^{-1}}X_{-}^{\bot }+\Pi _{%
\mathfrak{g}_{-}^{\bot }}\mathrm{ad}_{X_{-}^{g_{+}}}^{\tau }X_{+}^{\bot
}\right)%
\end{array}%
\right.
\end{equation*}%
which are relevant for the explicit form of the crossed adjoint actions

\begin{equation*}
\left\{ 
\begin{array}{l}
\mathrm{Ad}_{\left( g_{+},X_{+}^{\bot }\right) ^{-1}}^{H}\left(
X_{-},Y_{-}^{\bot }\right) =\left( g_{+},X_{+}^{\bot }\right) ^{-1}\left(
g_{+},X_{+}^{\bot }\right) ^{\left( X_{-},Y_{-}^{\bot }\right) } \\ 
\qquad \qquad \qquad \qquad \qquad +\left( X_{-},Y_{-}^{\bot }\right)
^{\left( g_{+},X_{+}^{\bot }\right) } \\ 
\\ 
\mathrm{Ad}_{\left( g_{-},X_{-}^{\bot }\right) }^{H}\left( X_{+},Y_{+}^{\bot
}\right) =\left( g_{-},X_{-}^{\bot }\right) ^{\left( X_{+},Y_{+}^{\bot
}\right) }\left( g_{-},X_{-}^{\bot }\right) ^{-1} \\ 
\qquad \qquad \qquad \qquad \qquad +\left( X_{+},Y_{+}^{\bot }\right)
^{\left( g_{-},X_{-}^{\bot }\right) }%
\end{array}%
\right.
\end{equation*}%
that are equivalent to 
\begin{equation}
\left\{ 
\begin{array}{l}
\mathrm{Ad}_{\left( g_{+},X_{+}^{\bot }\right) ^{-1}}^{H}\left(
X_{-},Y_{-}^{\bot }\right) =\left( \mathrm{Ad}_{g_{+}^{-1}}^{G}X_{-},\mathrm{%
ad}_{\mathrm{Ad}_{g_{+}^{-1}}^{G}X_{-}}^{\tau }X_{+}^{\bot }+\tau
_{g_{+}^{-1}}Y_{-}^{\bot }\right) \\ 
\\ 
\mathrm{Ad}_{\left( g_{-},X_{-}^{\bot }\right) }^{H}\left( X_{+},Y_{+}^{\bot
}\right) =\left( \mathrm{Ad}_{g_{-}}^{G}X_{+},-\mathrm{ad}_{\mathrm{Ad}%
_{g_{-}}^{G}X_{+}}^{\tau }\tau _{g_{-}}X_{-}^{\bot }+\tau
_{g_{-}}Y_{+}^{\bot }\right)%
\end{array}%
\right.  \label{crossed adj actions}
\end{equation}%
With this expressions we are ready to get the Poisson-Lie bivector on $%
H_{\pm }$.

\subsection{Lie bialgebra and Poisson-Lie structures}

Let us now work out the Lie bialgebra structures on $\mathfrak{h}_{_{\pm }}=%
\mathfrak{g}_{\pm }\oplus \mathfrak{g}_{\pm }^{\bot }$, and the associated
Poisson-Lie groups ones on $H_{_{\pm }}=G_{\pm }\oplus \mathfrak{g}_{\pm
}^{\bot }$. The Lie bracket on these semidirect sums were defined in $\left( %
\ref{Lie subalg}\right) $ and, in order to define the \emph{Lie cobracket} $%
\delta :\mathfrak{h}_{_{\pm }}\longrightarrow \mathfrak{h}_{_{\pm }}\otimes 
\mathfrak{h}_{_{\pm }}$, we introduce a bilinear form on $\mathfrak{h}%
\otimes \mathfrak{h}$ from the bilinear form $\left( \ref{Lie alg g+g 2}%
\right) $ as%
\begin{equation*}
\left( X\otimes Y,U\otimes V\right) _{\mathfrak{h}\otimes \mathfrak{h}%
}:=\left( X,U\right) _{\mathfrak{h}}\left( Y,V\right) _{\mathfrak{h}}
\end{equation*}%
$\forall X,Y,U,V\in \mathfrak{h}$. Thus, the Lie cobracket $\delta $ arises
from the relation 
\begin{eqnarray*}
&&\left( \left( X_{\mp }^{\prime },Y_{\mp }^{\prime \bot }\right) \otimes
\left( X_{\mp }^{\prime \prime },Y_{\mp }^{\prime \prime \bot }\right)
,\delta \left( X_{\pm },Y_{\pm }^{\bot }\right) \right) _{\mathfrak{h}%
\otimes \mathfrak{h}} \\
&=&\left( \left[ \left( X_{\mp }^{\prime },Y_{\mp }^{\prime \bot }\right)
,\left( X_{\mp }^{\prime \prime },Y_{\mp }^{\prime \prime \bot }\right) %
\right] ,\left( X_{\pm },Y_{\pm }^{\bot }\right) \right) _{\mathfrak{h}}
\end{eqnarray*}%
that in $2\times 2$ block matrix form means 
\begin{equation*}
\delta \left( X_{\pm },Y_{\pm }^{\bot }\right) =\left( 
\begin{array}{cc}
\delta \left( Y_{\pm }^{\bot }\right) & \tau ^{\ast }\left( X_{\pm }\right)
\\ 
-\tau ^{\ast }\left( X_{\pm }\right) & 0%
\end{array}%
\right)
\end{equation*}%
where $\delta :\mathfrak{g}_{\pm }^{\bot }\longrightarrow \mathfrak{g}_{\pm
}^{\bot }\otimes \mathfrak{g}_{\pm }^{\bot }$ such that%
\begin{equation*}
\left( X_{\mp }^{\prime }\otimes X_{\mp }^{\prime \prime },\delta \left(
Y_{\pm }^{\bot }\right) \right) _{\mathfrak{g}}:=\left( \left[ X_{\mp
}^{\prime },X_{\mp }^{\prime \prime }\right] ,Y_{\pm }^{\bot }\right) _{%
\mathfrak{g}}
\end{equation*}%
and $\tau ^{\ast }:\mathfrak{g}_{\pm }\longrightarrow \mathfrak{g}_{\pm
}^{\bot }\otimes \mathfrak{g}_{\pm }$ is%
\begin{equation*}
\left( X_{\mp }^{\prime }\otimes Y_{\mp }^{\prime \prime \bot },\tau ^{\ast
}\left( X_{\pm }\right) \right) _{\mathfrak{g}}:=\left( \mathrm{ad}_{X_{\mp
}^{\prime }}^{\tau }Y_{\mp }^{\prime \prime \bot },X_{\pm }\right) _{%
\mathfrak{g}}
\end{equation*}

\subsubsection{\textbf{The Poisson-Lie bivector on} $\mathit{H}_{\mathit{+}}$%
}

The Poisson-Lie bivector $\pi _{+}\in \Gamma \left( T^{\otimes
2}H_{+}\right) $, the sections of the vector bundle $T^{\otimes
2}H_{+}\longrightarrow H_{+}$, is defined by the relation 
\begin{eqnarray}
&&\left\langle \gamma \left( X_{-}^{\prime },Y_{-}^{\prime \bot }\right)
\left( g_{+},Y_{+}^{\bot }\right) ^{-1}\otimes \gamma \left( X_{-}^{\prime
\prime },Y_{-}^{\prime \prime \bot }\right) \left( g_{+},Y_{+}^{\bot
}\right) ^{-1},\pi _{+}\left( g_{+},Z_{+}^{\bot }\right) \right\rangle 
\notag \\
&=&\left( \Pi _{-}\mathrm{Ad}_{\left( g_{+},Z_{+}^{\bot }\right)
^{-1}}^{H}\left( X_{-}^{\prime },Y_{-}^{\prime \bot }\right) ,\Pi _{+}%
\mathrm{Ad}_{\left( g_{+},Z_{+}^{\bot }\right) ^{-1}}^{H}\left(
X_{-}^{\prime \prime },Y_{-}^{\prime \prime \bot }\right) \right) _{%
\mathfrak{h}}  \label{PL bivector h+ 1}
\end{eqnarray}%
In order to simplify the notation we introduce the projectors $\mathbb{A}%
_{\pm }^{G}\left( g\right) $, defined as%
\begin{equation*}
\mathbb{A}_{\pm }^{G}\left( g\right) :=\mathrm{Ad}_{g^{-1}}^{G}\Pi _{%
\mathfrak{g}_{\pm }}\mathrm{Ad}_{g}^{G}
\end{equation*}%
such that 
\begin{equation*}
\left\{ 
\begin{array}{l}
\mathbb{A}_{\pm }^{G}\left( g\right) \mathbb{A}_{\pm }^{G}\left( g\right) =%
\mathbb{A}_{\pm }^{G}\left( g\right) \\ 
\mathbb{A}_{\mp }^{G}\left( g\right) \mathbb{A}_{\pm }^{G}\left( g\right) =0
\\ 
\mathbb{A}_{+}^{G}\left( g\right) +\mathbb{A}_{-}^{G}\left( g\right) =Id%
\end{array}%
\right.
\end{equation*}%
which will be used in the following.

By using the expressions obtained in eq. $\left( \ref{crossed adj actions}%
\right) $, it takes the explicit form 
\begin{eqnarray}
&&\left\langle \gamma \left( X_{-}^{\prime },Y_{-}^{\prime \bot }\right)
\left( g_{+},Z_{+}^{\bot }\right) ^{-1}\otimes \gamma \left( X_{-}^{\prime
\prime },Y_{-}^{\prime \prime \bot }\right) \left( g_{+},Z_{+}^{\bot
}\right) ^{-1},\pi _{+}\left( g_{+},Z_{+}^{\bot }\right) \right\rangle
\label{PL bivector h+ 2} \\
&=&\left( \left( X_{-}^{\prime },Y_{-}^{\prime \bot }\right) ,\left( -%
\mathbb{A}_{-}^{G}\left( g_{+}^{-1}\right) X_{-}^{\prime \prime },\tau
_{g_{+}}\Pi _{\mathfrak{g}_{+}^{\bot }}\left( \tau
_{g_{+}^{-1}}Y_{-}^{\prime \prime \bot }+\mathrm{ad}_{\Pi _{\mathfrak{g}_{-}}%
\mathrm{Ad}_{g_{+}^{-1}}^{G}X_{-}^{\prime \prime }}^{\tau }Z_{+}^{\bot
}\right) \right) \right) _{\mathfrak{h}}  \notag
\end{eqnarray}%
Introducing the linear operator $\pi _{+\left( g_{+},Z_{+}^{\bot }\right)
}^{R}:\mathfrak{h}_{-}\longrightarrow \mathfrak{h}_{+}$ such that%
\begin{eqnarray*}
&&\left\langle \gamma \left( X_{-}^{\prime },Y_{-}^{\prime \bot }\right)
\left( g_{+},Z_{+}^{\bot }\right) ^{-1}\otimes \gamma \left( X_{-}^{\prime
\prime },Y_{-}^{\prime \prime \bot }\right) \left( g_{+},Z_{+}^{\bot
}\right) ^{-1},\pi _{+}\left( g_{+},Z_{+}^{\bot }\right) \right\rangle \\
&=&\left( \left( X_{-}^{\prime },Y_{-}^{\prime \bot }\right) ,\pi _{+\left(
g_{+},Z_{+}^{\bot }\right) }^{R}\left( X_{-}^{\prime \prime },Y_{-}^{\prime
\prime \bot }\right) \right) _{\mathfrak{h}}
\end{eqnarray*}%
we get, in components in $\mathfrak{g}_{+}\oplus \mathfrak{g}_{+}^{\bot }$,
the operator block matrix 
\begin{equation*}
\pi _{+\left( g_{+},Z_{+}^{\bot }\right) }^{R}\left( 
\begin{array}{c}
X_{-}^{\prime \prime } \\ 
Y_{-}^{\prime \prime \bot }%
\end{array}%
\right) =\left( 
\begin{array}{cc}
\mathbb{A}_{+}^{G}\left( g_{+}^{-1}\right) & 0 \\ 
\tau _{g_{+}}\Pi _{+}^{\bot }\tilde{\varphi}\left( Z_{+}^{\bot }\right) \Pi
_{-}\mathrm{Ad}_{g_{+}^{-1}}^{G} & \tau _{g_{+}}\Pi _{+}^{\bot }\tau
_{g_{+}^{-1}}%
\end{array}%
\right) \left( 
\begin{array}{c}
X_{-}^{\prime \prime } \\ 
Y_{-}^{\prime \prime \bot }%
\end{array}%
\right)
\end{equation*}%
Here we introduced the map $\tilde{\varphi}:\mathfrak{g}\longrightarrow
End_{vec}\left( \mathfrak{g}\right) $ such that%
\begin{equation}
\tilde{\varphi}\left( Z_{+}^{\bot }\right) X_{-}^{\prime \prime }:=\mathrm{ad%
}_{X_{-}^{\prime \prime }}^{\tau }Z_{+}^{\bot }  \label{phi orlado}
\end{equation}

For a couple of functions $\mathcal{F},\mathcal{H}$ on $H_{+}=G_{+}\ltimes 
\mathfrak{g}_{+}^{\bot }$ and the definition of the Poisson bracket in terms
of the bivector $\pi _{H}^{+}$%
\begin{equation*}
\left\{ \mathcal{F},\mathcal{H}\right\} _{PL}\left( g_{+},Z_{+}^{\bot
}\right) =\left\langle d\mathcal{F}\wedge d\mathcal{H},\pi _{+}\left(
g_{+},Z_{+}^{\bot }\right) \right\rangle
\end{equation*}%
we use the expression $\left( \ref{PL bivector h+ 2}\right) $ to obtain 
\begin{eqnarray*}
&&\left\{ \mathcal{F},\mathcal{H}\right\} _{PL}\left( g_{+},Z_{+}^{\bot
}\right) \\
&=&\left\langle g_{+}\mathbf{d}\mathcal{H},\bar{\psi}\left( \delta \mathcal{F%
}\right) \right\rangle -\left\langle g_{+}\mathbf{d}\mathcal{F},\bar{\psi}%
\left( \delta \mathcal{H}\right) \right\rangle +\left\langle \psi \left(
Z_{+}^{\bot }\right) ,\left[ \bar{\psi}\left( \delta \mathcal{F}\right) ,%
\bar{\psi}\left( \delta \mathcal{H}\right) \right] \right\rangle
\end{eqnarray*}%
for $d\mathcal{F}=\left( \mathbf{d}\mathcal{F},\delta \mathcal{F}\right) \in
T^{\ast }H_{+}$.

Also, from this Poisson-Lie bivector we can retrieve the infinitesimal
generator of the dressing action $\left( \ref{inf gen dres transad}\right) $
by the relation%
\begin{equation}
\left( g_{+},Z_{+}^{\bot }\right) ^{\left( X_{-},Y_{-}^{\bot }\right)
}=\left( R_{\left( g_{+},Z_{+}^{\bot }\right) }\right) _{\ast }\left( \pi
_{+\left( g_{+},Z_{+}^{\bot }\right) }^{R}\left( X_{-},Y_{-}^{\bot }\right)
\right)  \label{PL bivector h+ 7}
\end{equation}%
It is worth to recall that the symplectic leaves of the Poisson-Lie
structure are the orbits of the dressing actions, the integral submanifolds
of the dressing vector fields, and $\left( e,0\right) \in H_{+}$ is a
1-point orbit.

\subsubsection{\textbf{The Poisson-Lie bivector on} $\mathit{H}_{\mathit{-}}$%
.}

Analogously to the previous definition, the Poisson-Lie bivector $\pi
_{-}\in \Gamma \left( T^{\otimes 2}H_{-}\right) $, is defined by the relation%
\begin{eqnarray}
&&\left\langle \left( g_{-},Z_{-}^{\bot }\right) ^{-1}\gamma \left(
X_{+}^{\prime },Y_{+}^{\prime \bot }\right) \otimes \left( g_{-},Z_{-}^{\bot
}\right) ^{-1}\gamma \left( X_{+}^{\prime \prime },Y_{+}^{\prime \prime \bot
}\right) ,\pi _{-}\left( g_{-},Z_{-}^{\bot }\right) \right\rangle  \notag \\
&=&\left( \Pi _{-}\mathrm{Ad}_{\left( g_{-},Z_{-}^{\bot }\right) }^{H}\left(
X_{+}^{\prime },Y_{+}^{\prime \bot }\right) ,\Pi _{+}\mathrm{Ad}_{\left(
g_{-},Z_{-}^{\bot }\right) }^{H}\left( X_{+}^{\prime \prime },Y_{+}^{\prime
\prime \bot }\right) \right) _{\mathfrak{h}}  \label{PL bivector h- 1}
\end{eqnarray}%
From the relation $\left( \ref{crossed adj actions}\right) $, we get 
\begin{eqnarray*}
&&\left\langle \left( g_{-},Z_{-}^{\bot }\right) ^{-1}\gamma \left(
X_{+}^{\prime },Y_{+}^{\prime \bot }\right) \otimes \left( g_{-},Z_{-}^{\bot
}\right) ^{-1}\gamma \left( X_{+}^{\prime \prime },Y_{+}^{\prime \prime \bot
}\right) ,\pi _{-}\left( g_{-},Z_{-}^{\bot }\right) \right\rangle \\
&=&\left( \mathbb{A}_{+}^{G}\left( g_{-}\right) X_{+}^{\prime
},Y_{+}^{\prime \prime \bot }\right) _{\mathfrak{g}}+\left( Y_{+}^{\prime
\bot },\mathbb{A}_{-}^{G}\left( g_{-}\right) X_{+}^{\prime \prime }\right) _{%
\mathfrak{g}}-\left( Z_{-}^{\bot },\left[ X_{+}^{\prime },X_{+}^{\prime
\prime }\right] \right) _{\mathfrak{g}} \\
&&+\left( Z_{-}^{\bot },\left[ X_{+}^{\prime },\mathbb{A}_{-}^{G}\left(
g_{-}\right) X_{+}^{\prime \prime }\right] \right) _{\mathfrak{g}}+\left(
Z_{-}^{\bot },\left[ \mathbb{A}_{-}^{G}\left( g_{-}\right) X_{+}^{\prime
},X_{+}^{\prime \prime }\right] \right) _{\mathfrak{g}}
\end{eqnarray*}%
The PL bracket for functions $\mathcal{F},\mathcal{H}$ on $%
H_{-}=G_{-}\ltimes \mathfrak{g}_{-}^{\bot }$ is obtained by identifying the
differential $d\mathcal{F},d\mathcal{H}$ with $\left( g_{-},Z_{-}^{\bot
}\right) ^{-1}\gamma \left( X_{+}^{\prime },Y_{+}^{\prime \bot }\right) $, $%
\left( g_{-},Z_{-}^{\bot }\right) ^{-1}\gamma \left( X_{+}^{\prime \prime
},Y_{+}^{\prime \prime \bot }\right) $ respectively. Then, making explicit
the left translation we obtain%
\begin{eqnarray*}
\left\{ \mathcal{F},\mathcal{H}\right\} _{PL}\left( g_{-},Z_{-}^{\bot
}\right) &=&-\left\langle g_{-}\mathbf{d}\mathcal{H},\mathbb{A}%
_{-}^{G}\left( g_{-}\right) \bar{\psi}\left( \delta \mathcal{F}\right)
\right\rangle \\
&&+\left\langle g_{-}\mathbf{d}\mathcal{F},\mathbb{A}_{-}^{G}\left(
g_{-}\right) \bar{\psi}\left( \delta \mathcal{H}\right) \right\rangle \\
&&-\left\langle \psi \left( Z_{-}^{\bot }\right) ,\left[ \bar{\psi}\left(
\delta \mathcal{F}\right) ,\bar{\psi}\left( \delta \mathcal{H}\right) \right]
\right\rangle
\end{eqnarray*}%
Alternatively, we may write the PL bivector in terms of the associated
linear operator $\pi _{-\left( h_{-},Z_{-}^{\bot }\right) }^{L}:\mathfrak{h}%
_{-}^{\ast }\longrightarrow \mathfrak{h}_{-}$ defined from%
\begin{eqnarray*}
&&\left\langle \left( g_{-},Z_{-}^{\bot }\right) ^{-1}\gamma \left(
X_{+}^{\prime },Y_{+}^{\prime \bot }\right) \otimes \left( g_{-},Z_{-}^{\bot
}\right) ^{-1}\gamma \left( X_{+}^{\prime \prime },Y_{+}^{\prime \prime \bot
}\right) ,\pi _{-}\left( g_{-},Z_{-}^{\bot }\right) \right\rangle \\
&=&\left( \left( X_{+}^{\prime },Y_{+}^{\prime \bot }\right) ,\pi _{-\left(
g_{-},Z_{-}^{\bot }\right) }^{L}\left( X_{+}^{\prime \prime },Y_{+}^{\prime
\prime \bot }\right) \right) _{\mathfrak{h}}
\end{eqnarray*}%
we get, in terms of components in the direct sum, that%
\begin{eqnarray*}
&&\medskip \pi _{-\left( g_{-},Z_{-}^{\bot }\right) }^{L}\left( \gamma
\left( 
\begin{array}{c}
X_{+}^{\prime \prime } \\ 
Y_{+}^{\prime \prime \bot }%
\end{array}%
\right) \right) \\
&=&\left( 
\begin{array}{cc}
\Pi _{-}\left( \mathbb{A}_{+}^{G}\left( g_{-}\right) \right) & 0 \\ 
\tilde{\varphi}\left( Z_{-}^{\bot }\right) \mathbb{A}_{+}^{G}\left(
g_{-}\right) -\tau _{g_{-}^{-1}}\Pi _{+}^{\bot }\tau _{g_{-}}\tilde{\varphi}%
\left( Z_{-}^{\bot }\right) & \tau _{g_{+}}\Pi _{+}^{\bot }\tau _{g_{+}^{-1}}%
\end{array}%
\right) \left( 
\begin{array}{c}
X_{+}^{\prime \prime } \\ 
Y_{+}^{\prime \prime \bot }%
\end{array}%
\right)
\end{eqnarray*}%
where we used $\tilde{\varphi}:\mathfrak{g}\longrightarrow End_{vec}\left( 
\mathfrak{g}\right) $ introduced in $\left( \ref{phi orlado}\right) $, such
that%
\begin{equation*}
\tilde{\varphi}\left( Z_{-}^{\bot }\right) X_{+}:=\mathrm{ad}_{X_{+}}^{\tau
}Z_{-}^{\bot }
\end{equation*}%
The infinitesimal generator of the dressing action are then%
\begin{equation}
\left( g_{-},Z_{-}^{\bot }\right) ^{\left( X_{+},Y_{+}^{\bot }\right)
}=\left( \pi _{-\left( e,0\right) }^{L}\left( \gamma \left(
X_{+},Y_{+}^{\bot }\right) \right) \right) \left( g_{-},Z_{-}^{\bot }\right)
\label{PL bivector h- 2}
\end{equation}%
retrieving the result obtained in $\left( \ref{inf gen dres transad}\right) $%
.

\section{Integrability on $\mathfrak{g}$ from AKS on $\mathfrak{h}$}

Despite AKS integrable system theory also works in absence of an
Ad-invariant bilinear form \cite{Ovando 1},\cite{Ovando 2}, the $\mathrm{Ad}%
^{H}$-invariant bilinear form $\left( \ref{Lie alg g+g 2}\right) $ on the
semidirect product $\mathfrak{h}=\mathfrak{g}\ltimes _{\tau }\mathfrak{g}$
brings back to the standard framework of AKS theory. We now briefly review
it in the context of the semidirect product $\mathfrak{g}\ltimes _{\tau }%
\mathfrak{g}$.

Let us consider now $\mathfrak{h}^{\ast }$ equipped with the Lie-Poisson
structure. Then, turning the linear isomorphism $\mathfrak{h}\overset{\gamma 
}{\longrightarrow }\mathfrak{h}^{\ast }$ into a Poisson map, we define a
Poisson structure on $\mathfrak{h}$ such that for any couple of functions $%
\mathsf{F},\mathsf{H}:\mathfrak{h}\longrightarrow \mathbb{R}$, it is 
\begin{equation*}
\left\{ \mathsf{F},\mathsf{H}\right\} _{\mathfrak{h}}\left( X\right)
:=\left\{ \mathsf{F}\circ \gamma ^{-1},\mathsf{H}\circ \gamma ^{-1}\right\}
_{\mathfrak{h}^{\ast }}\left( \psi \left( X\right) \right) =\left( X,\left[ 
\mathfrak{L}_{\mathsf{F}}(X),\mathfrak{L}_{\mathsf{H}}(X)\right] \right) _{%
\mathfrak{h}}
\end{equation*}%
where the Legendre transform $\mathfrak{L}_{\mathsf{F}}:\mathfrak{h}%
\longrightarrow \mathfrak{h}$ is defined as%
\begin{equation*}
\left( \mathfrak{L}_{\mathsf{F}}\left( X\right) ,Y\right) _{\mathfrak{h}%
}=\left. \frac{d\mathsf{F}\left( X+tY\right) )}{dt}\right\vert _{t=0}
\end{equation*}

Let us now consider a function $\mathsf{F}:\mathfrak{h}\longrightarrow 
\mathbb{R}$, and let $\mathcal{F}:=\mathsf{F}\circ \imath _{K_{-}}$ with $%
\imath _{K_{-}}:\mathfrak{h}_{+}\longrightarrow \mathfrak{h}$, for some $%
K_{-}\in \mathfrak{h}_{-}$, such that%
\begin{equation*}
\imath _{K_{-}}(X_{+})=X_{+}+K_{-}
\end{equation*}%
The Legendre transform $\mathfrak{L}_{\mathcal{F}}:\mathfrak{h}%
_{+}\longrightarrow \mathfrak{h}_{-}$ relates with $\mathfrak{L}_{\mathsf{F}%
} $ as 
\begin{equation*}
\mathfrak{L}_{\mathcal{F}}(X_{+})=\Pi _{\mathfrak{g}_{-}}\mathfrak{L}_{%
\mathsf{F}}(\imath \left( X_{+}+K_{-}\right) ).
\end{equation*}
Therefore, the Lie-Poisson bracket on $\mathfrak{h}_{+}$ of $\mathcal{F},%
\mathcal{H}:\mathfrak{h}_{+}\longrightarrow \mathbb{R}$ coincides with that
of $\mathsf{F},\mathsf{H}:\mathfrak{h}\longrightarrow \mathbb{R}$ restricted
to $\mathfrak{h}_{+}$ 
\begin{eqnarray*}
\left\{ \mathcal{F},\mathcal{H}\right\} _{\mathfrak{h}_{+}}\left(
X_{+}\right) &=&\left( X_{+},\left[ \mathfrak{L}_{\mathcal{F}}(X_{+}),%
\mathfrak{L}_{\mathcal{H}}(X_{+})\right] _{-}\right) _{\mathfrak{h}} \\
&=&\left\langle d\mathsf{F},\Pi _{\mathfrak{h}_{+}}\mathrm{ad}_{\Pi _{%
\mathfrak{h}_{-}}\mathfrak{L}_{\mathsf{H}}\left( X_{+}+K_{-}\right) }^{%
\mathfrak{h}}\left( X_{+}+K_{-}\right) \right\rangle
\end{eqnarray*}%
where $\Pi _{\mathfrak{h}_{+}}\mathrm{ad}_{\Pi _{\mathfrak{h}_{-}}\mathfrak{L%
}_{\mathsf{H}}\left( X_{+}+K_{-}\right) }^{\mathfrak{h}}\left(
X_{+}+K_{-}\right) $ is the projection of the hamiltonian vector field of $%
\mathsf{H}$ on $\mathfrak{h}_{+}$. Then 
\begin{eqnarray}
\left\{ \mathcal{F},\mathcal{H}\right\} _{\mathfrak{h}_{+}}\left(
X_{+}\right) &=&\left. \left\{ \mathsf{F},\mathsf{H}\right\} _{\mathfrak{h}%
}\right\vert _{+}\left( X_{+}+K_{-}\right)  \label{restr pb h+} \\
&=&\left( X_{+},\left[ \Pi _{\mathfrak{h}_{-}}\mathfrak{L}_{\mathsf{F}%
}(X_{+}+K_{-}),\Pi _{\mathfrak{h}_{-}}\mathfrak{L}_{\mathsf{H}}\left(
X_{+}+K_{-}\right) \right] _{-}\right) _{\mathfrak{h}}  \notag
\end{eqnarray}

Let $\mathsf{F}$ be\textit{\ }an $\mathrm{Ad}^{H}$-invariant function on $%
\mathfrak{h}$, then the following relations holds: $\forall h\in G$, \textit{%
\ }%
\begin{equation*}
\mathrm{Ad}_{h}^{H}\mathfrak{L}_{\mathsf{F}}(X)=\mathfrak{L}_{\mathsf{F}}(%
\mathrm{Ad}_{h}^{H}X)
\end{equation*}%
and $\forall X\in \mathfrak{h}$,%
\begin{equation*}
\begin{array}{ccc}
\mathrm{ad}_{\mathfrak{L}_{\mathsf{F}}(X)}^{\mathfrak{h}}X=0 & 
\Longrightarrow & \mathrm{ad}_{\Pi _{\mathfrak{h}_{-}}\mathfrak{L}_{\mathsf{F%
}}(X)}^{\mathfrak{h}}X=-\mathrm{ad}_{\Pi _{\mathfrak{h}_{+}}\mathfrak{L}_{%
\mathsf{F}}(X)}^{\mathfrak{h}}X%
\end{array}%
\end{equation*}

By using this results, it is easy to show the following statement.

\begin{description}
\item[AKS\ Theorem:] \textit{Let} $\gamma \left( K_{-}\right) $\textit{\ be
a character of} $\mathfrak{h}_{+}$. \textit{Then, the restriction to the
immersed submanifold }$\imath _{K_{-}}:\mathfrak{h}_{+}\longrightarrow 
\mathfrak{h}$\textit{\ of }$\mathrm{Ad}^{H}$\textit{-invariant functions on }%
$\mathfrak{h}$ \textit{gives rise to a nontrivial abelian Poisson} \textit{%
algebra.}

\item[Remark:] \textit{The condition} $\gamma \left( K_{-}\right) \in 
\mathrm{char~}\mathfrak{h}_{+}$ \textit{means, }$\forall X_{+}\in \mathfrak{h%
}_{+}$,%
\begin{equation*}
ad_{X_{+}}^{\ast }\gamma \left( K_{-}\right) =0\Longleftrightarrow \Pi _{%
\mathfrak{h}_{-}}\mathrm{ad}_{X_{+}}^{\mathfrak{h}}K_{-}=0
\end{equation*}%
\textit{This condition in terms of components in} $\mathfrak{g}\oplus 
\mathfrak{g}$\textit{\ implies} 
\begin{equation*}
\begin{cases}
\Pi _{\mathfrak{h}_{-}}\left[ \Pi _{1}X_{+},\Pi _{1}K_{-}\right] _{\mathfrak{%
g}}=0 \\ 
\Pi _{\mathfrak{h}_{-}}^{\bot }\mathrm{ad}_{\Pi _{1}X_{+}}^{\tau }\Pi
_{2}K_{-}=0%
\end{cases}%
\end{equation*}%
\textit{where }$\Pi _{i}$ , $i=1,2$\textit{, stand for the proyectors on the
first and second component in the semidirect sum }$\mathfrak{g}\ltimes 
\mathfrak{g}$\textit{, respectively}.
\end{description}

The dynamics on these submanifolds is ruled by the hamiltonian vector fields
arising from the process of restriction, as described in the following
proposition.

\begin{description}
\item[Proposition:] \textit{The hamiltonian vector field associated to the
Poisson bracket on }$\mathfrak{h}_{+}$\textit{, eq. }$\left( \ref{restr pb
h+}\right) $\textit{,} \textit{is}%
\begin{eqnarray*}
V_{\mathsf{H}}\left( X_{+}+K_{-}\right) &=&\Pi _{\mathfrak{h}_{+}}\mathrm{ad}%
_{\Pi _{\mathfrak{h}_{-}}\mathfrak{L}_{\mathsf{H}}\left( X_{+}+K_{-}\right)
}^{\mathfrak{h}}\left( X_{+}+K_{-}\right) \\
&=&-\mathrm{ad}_{\Pi _{\mathfrak{h}_{+}}\mathfrak{L}_{\mathsf{F}%
}(X_{+}+K_{-})}^{\mathfrak{h}}\left( X_{+}+K_{-}\right)
\end{eqnarray*}%
\textit{for }$X_{+}^{\bot }\in \mathfrak{h}_{+}^{\bot }$.
\end{description}

The integral curves of this hamiltonian vector field, a \emph{dressing
vector field indeed}, are the orbits of a particular curve in $H$, as
explained in the following proposition.

\begin{description}
\item[Propositon:] \textit{The Hamilton equation of motion} on $\imath
_{K_{-}}\left( \mathfrak{h}_{+}\right) \subset \mathfrak{h}$ \textit{are}%
\begin{equation}
\begin{cases}
\dot{Z}\left( t\right) =V_{\mathsf{H}}\left( Z\left( t\right) \right) \\ 
Z_{\circ }=Z\left( 0\right) =X_{+\circ }+K_{-}%
\end{cases}
\label{Ham sist on h+}
\end{equation}%
\textit{with }$Z\left( t\right) =X_{+}\left( t\right) +K_{-}$\textit{, and} $%
\mathsf{H}$ \textit{is an} $\mathrm{Ad}^{H}$\textit{-invariant function. It
is solved by factorization: if} $h_{+},h_{-}:\mathbb{R}\rightarrow H_{\pm }$ 
\textit{are curves on these groups defined by}%
\begin{equation*}
\exp \left( t\mathfrak{L}_{\mathsf{H}}(Z_{\circ })\right) =h_{+}\left(
t\right) h_{-}\left( t\right)
\end{equation*}%
\textit{the solution of the above Hamilton equation is }%
\begin{equation*}
Z\left( t\right) =\mathrm{Ad}_{h_{+}^{-1}\left( t\right) }^{H}\left(
Z_{\circ }\right)
\end{equation*}
\end{description}

\textbf{Proof: }The Hamilton equation of motion are%
\begin{equation}
\dot{Z}\left( t\right) =-\mathrm{ad}_{\Pi _{\mathfrak{h}_{+}}\mathfrak{L}_{%
\mathsf{H}}\left( Z\left( t\right) \right) }^{\mathfrak{h}}Z\left( t\right)
\label{Eq h+}
\end{equation}%
It is solved by 
\begin{equation*}
Z\left( t\right) =\mathrm{Ad}_{h_{+}^{-1}\left( t\right) }^{H}\left(
Z_{\circ }\right) \in H_{+}
\end{equation*}%
with the curve $t\longmapsto h_{+}\left( t\right) \subset H_{+}$ solving 
\begin{equation}
h_{+}^{-1}\left( t\right) \dot{h}_{+}\left( t\right) =\Pi _{\mathfrak{h}_{+}}%
\mathfrak{L}_{\mathsf{H}}(Z\left( t\right) )  \label{Eq H+}
\end{equation}

Now, let us consider a curve $t\longmapsto h\left( t\right) =e^{t\mathfrak{L}%
_{\mathsf{H}}(X_{+\circ }+K_{-})}\subset H$, for constant $X_{+\circ }+K_{-}$%
, which solves the differential equation%
\begin{equation*}
\dot{h}\left( t\right) h^{-1}\left( t\right) =\mathfrak{L}_{\mathsf{H}%
}(Z_{\circ })
\end{equation*}%
so, as $H=H_{+}H_{-}$ and $\mathfrak{h}=\mathfrak{h}_{+}\oplus \mathfrak{h}%
_{-}$ we have $h\left( t\right) =h_{+}\left( t\right) h_{-}\left( t\right) $
with $h_{+}\left( t\right) \in H_{+}$ and $h_{-}\left( t\right) \in H_{-}$,
hence $\dot{h}h^{-1}=\dot{h}_{+}h_{+}^{-1}+\mathrm{Ad}_{h_{+}}^{H}\dot{h}%
_{-}h_{-}^{-1}$ and, since $\mathsf{H}$ is $\mathrm{Ad}^{H}$-invariant, the
equation of motion turns in 
\begin{equation*}
h_{+}^{-1}\dot{h}_{+}+\dot{h}_{-}h_{-}^{-1}=\mathfrak{L}_{\mathsf{H}}(%
\mathrm{Ad}_{h_{+}^{-1}}^{H}\left( Z_{\circ }\right) )
\end{equation*}%
from where we conclude that%
\begin{equation*}
\left\{ 
\begin{array}{c}
\medskip h_{+}^{-1}\left( t\right) \dot{h}_{+}\left( t\right) =\Pi _{%
\mathfrak{h}_{+}}\mathfrak{L}_{\mathsf{H}}(Z\left( t\right) ) \\ 
\dot{h}_{-}\left( t\right) h_{-}^{-1}\left( t\right) =\Pi _{\mathfrak{h}_{-}}%
\mathfrak{L}_{\mathsf{H}}(Z\left( t\right) )%
\end{array}%
\right.
\end{equation*}%
Here we can see that the first equation coincides with $\left( \ref{Eq H+}%
\right) $, thus showing that the factor $h_{+}\left( t\right) $ of the
decomposition $e^{t\mathfrak{L}_{\mathsf{H}}(X_{+})}=h_{+}\left( t\right)
h_{-}\left( t\right) $ solves the hamiltonian system $\left( \ref{Eq h+}%
\right) $.$\blacksquare $

In order to write the Hamilton equation $\left( \ref{Ham sist on h+}\right) $
in terms of the components $\mathfrak{h}=\mathfrak{g}\oplus \mathfrak{g}$,
we write $Z\left( t\right) =X_{+}\left( t\right) +K_{-}$ as 
\begin{equation*}
Z\left( t\right) =\left( \Pi _{1}\left( X_{+}\left( t\right) +K_{-}\right)
,\Pi _{2}\left( X_{+}\left( t\right) +K_{-}\right) \right)
\end{equation*}%
Then, the evolution equations for each component are 
\begin{equation}
\begin{cases}
\Pi _{1}\dot{X}_{+}\left( t\right) =\left[ \Pi _{1}X_{+}\left( t\right) +\Pi
_{1}K_{-},\Pi _{1}\Pi _{\mathfrak{h}_{+}}\mathfrak{L}_{\mathsf{H}}\left(
Z\left( t\right) \right) \right] \\ 
\\ 
\Pi _{2}\dot{X}_{+}\left( t\right) =\mathrm{ad}_{\Pi _{1}X_{+}\left(
t\right) +\Pi _{1}K_{-}}^{\tau }\Pi _{2}\Pi _{\mathfrak{h}_{+}}\mathfrak{L}_{%
\mathsf{H}}\left( Z\left( t\right) \right) \\ 
\qquad \qquad \qquad -\mathrm{ad}_{\Pi _{1}\Pi _{\mathfrak{h}_{+}}\mathfrak{L%
}_{\mathsf{H}}\left( Z\left( t\right) \right) }^{\tau }\left( \Pi
_{2}X_{+}\left( t\right) +\Pi _{2}K_{-}\right)%
\end{cases}
\label{Z5}
\end{equation}%
whose solutions are obtained from the components of the curve 
\begin{equation*}
Z\left( t\right) =\mathrm{Ad}_{h_{+}^{-1}\left( t\right) }^{H}\left(
Z_{\circ }\right)
\end{equation*}%
Explicitly they are 
\begin{equation*}
\begin{cases}
\Pi _{1}X_{+}\left( t\right) =\mathrm{Ad}_{g_{+}^{-1}\left( t\right)
}^{G}\Pi _{1}\left( Z_{\circ }\right) \\ 
\\ 
\Pi _{2}X_{+}\left( t\right) =\tau _{g_{+}^{-1}\left( t\right) }\Pi
_{2}\left( Z_{\circ }\right) +\mathrm{ad}_{\mathrm{Ad}_{g_{+}^{-1}\left(
t\right) }^{G}\Pi _{1}\left( Z_{\circ }\right) }^{\tau }Y_{+}^{\bot }\left(
t\right)%
\end{cases}%
\end{equation*}%
where $\left( g_{+}\left( t\right) ,Y_{+}^{\bot }\left( t\right) \right) $
is the $H_{+}=\mathfrak{g}_{+}\oplus \mathfrak{g}_{+}^{\bot }$ factor of the
exponential curve in $H=G\ltimes \mathfrak{g}$, namely%
\begin{equation*}
\mathrm{Exp}^{\cdot }\left( t\mathfrak{L}_{\mathsf{H}}\left( Z_{\circ
}\right) \right) =\left( e^{t\Pi _{1}\mathfrak{L}_{\mathsf{H}}\left(
Z_{\circ }\right) },-\sum_{n=1}^{\infty }\frac{\left( -1\right) ^{n}}{n!}%
t^{n}\left( \mathrm{ad}_{\Pi _{1}\mathfrak{L}_{\mathsf{H}}\left( Z_{\circ
}\right) }^{\tau }\right) ^{n-1}\Pi _{2}\mathfrak{L}_{\mathsf{H}}\left(
Z_{\circ }\right) \right)
\end{equation*}%
where $\mathfrak{L}_{\mathsf{H}}\left( Z_{\circ }\right) =\left( \Pi _{1}%
\mathfrak{L}_{\mathsf{H}}\left( Z_{\circ }\right) ,\Pi _{2}\mathfrak{L}_{%
\mathsf{H}}\left( Z_{\circ }\right) \right) \in \mathfrak{g}\oplus \mathfrak{%
g}$.

Let us restrict to the subspace 
\begin{equation*}
\mathfrak{g}_{2}:=\left\{ 0\right\} \oplus \mathfrak{g}:=\left\{ \left(
0,Y\right) /Y\in \mathfrak{g}\right\}
\end{equation*}%
Then, the $\mathrm{Ad}^{H}$-invariant function $\mathsf{H}$ reduces to an $%
\tau ^{G}$-invariant function on $\mathfrak{g}_{2}$ since 
\begin{equation*}
\left. \mathrm{Ad}_{\left( g,Z\right) }^{H}\left( X,Y\right) \right\vert _{%
\mathfrak{g}_{2}}=\left( 0,\tau _{g}Y\right)
\end{equation*}%
then, if $\imath :\mathfrak{g}_{2}\longrightarrow \mathfrak{h}$ is the
injection then $\mathsf{h}:=\mathsf{H}\circ \imath :\mathfrak{g}%
_{2}\longrightarrow \mathbb{R}$ is the restriction of $\mathsf{H}$ to $%
\mathfrak{g}_{2}$. Then 
\begin{eqnarray*}
\left( \mathfrak{L}_{\mathsf{h}}\left( X\right) ,Y\right) _{\mathfrak{h}}
&=&\left. \frac{d\mathsf{H}\circ \imath \left( X+tY\right) )}{dt}\right\vert
_{t=0} \\
&=&\left( \mathfrak{L}_{\mathsf{H}}\imath \left( X\right) ,\left( 0,Y\right)
\right) _{\mathfrak{h}}=\left( \Pi _{1}\mathfrak{L}_{\mathsf{H}}\imath
\left( X\right) ,Y\right) _{\mathfrak{g}}
\end{eqnarray*}%
so, we conclude that%
\begin{equation*}
\Pi _{1}\mathfrak{L}_{\mathsf{H}}\imath \left( X\right) =\mathfrak{L}_{%
\mathsf{h}}\left( X\right)
\end{equation*}%
The restriction of the differential equations $\left( \ref{Z5}\right) $ to
this subspace are%
\begin{equation*}
\Pi _{2}\dot{X}_{+}\left( t\right) =-\mathrm{ad}_{\Pi _{\mathfrak{g}_{+}}%
\mathfrak{L}_{\mathsf{h}}\left( \Pi _{2}X_{+}\left( t\right) \right) }^{\tau
}\left( \Pi _{2}Z_{\circ }\right)
\end{equation*}%
that reproduces the equation integrable via AKS obtained in ref. \cite%
{Ovando 1},\cite{Ovando 2}.

\section{Examples of Lie algebras with no bi-invariant metrics}

\subsection{A three step nilpotent Lie algebra}

Let us consider the three step nilpotent Lie algebra $\mathfrak{g}$
generated by $\left\{ e_{1},e_{2},e_{3},e_{4}\right\} $, see ref. \cite%
{Ovando 1}, defined by the nonvanishing Lie brackets 
\begin{equation}
\begin{array}{ccc}
\left[ e_{4},e_{1}\right] =e_{2} & , & \left[ e_{4},e_{2}\right] =e_{3}%
\end{array}
\label{3step 0}
\end{equation}%
and the metric is determined by the nonvanishing pairings 
\begin{equation*}
\begin{array}{ccc}
\left( e_{2},e_{2}\right) _{\mathfrak{g}}=\left( e_{4},e_{4}\right) _{%
\mathfrak{g}}=1 & , & \left( e_{1},e_{3}\right) _{\mathfrak{g}}=-1%
\end{array}%
\end{equation*}%
It can be decomposed in two different direct sums $\mathfrak{g}=\mathfrak{g}%
_{+}\oplus \mathfrak{g}_{-}$ or $\mathfrak{g}=\mathfrak{g}_{+}^{\bot }\oplus 
\mathfrak{g}_{-}^{\bot }$ with%
\begin{equation*}
\begin{array}{lll}
\mathfrak{g}_{+}=lspan\left\{ e_{2},e_{3},e_{4}\right\} & , & \mathfrak{g}%
_{+}^{\bot }=lspan\left\{ e_{3}\right\} \\ 
\mathfrak{g}_{-}=lspan\left\{ e_{1}\right\} & , & \mathfrak{g}_{-}^{\bot
}=lspan\left\{ e_{1},e_{2},e_{4}\right\}%
\end{array}%
\end{equation*}%
where $\mathfrak{g}_{+}$ and $\mathfrak{g}_{-}$ are Lie subalgebras of $%
\mathfrak{g}$.

The nonvanishing $\tau $-action action of $\mathfrak{g}$ on itself are%
\begin{equation*}
\begin{array}{lllllll}
\mathrm{ad}_{e_{2}}^{\tau }e_{1}=-e_{4} & , & \mathrm{ad}_{e_{1}}^{\tau
}e_{2}=e_{4} & , & \mathrm{ad}_{e_{4}}^{\tau }e_{1}=e_{2} & , & \mathrm{ad}%
_{e_{4}}^{\tau }e_{2}=e_{3}%
\end{array}%
\end{equation*}%
and the Lie bracket in the semidirect sum Lie algebra $\mathfrak{h}=%
\mathfrak{g}\ltimes \mathfrak{g}$, see eq. $\left( \ref{Lie alg g+g 1}%
\right) $, is defined by the following non trivial ones%
\begin{equation*}
\begin{array}{ccc}
\left[ \left( e_{4},0\right) ,\left( e_{1},0\right) \right] =\left(
e_{2},0\right) &  & \left[ \left( e_{2},0\right) ,\left( 0,e_{1}\right) %
\right] =\left( 0,-e_{4}\right) \\ 
\left[ \left( e_{4},0\right) ,\left( e_{2},0\right) \right] =\left(
e_{3},0\right) &  & \left[ \left( e_{4},0\right) ,\left( 0,e_{2}\right) %
\right] =\left( 0,e_{3}\right) \\ 
\left[ \left( e_{1},0\right) ,\left( 0,e_{2}\right) \right] =\left(
0,e_{4}\right) &  & \left[ \left( e_{4},0\right) ,\left( 0,e_{1}\right) %
\right] =\left( 0,-e_{2}\right)%
\end{array}%
\end{equation*}

The Ad-invariant symmetric nondegenerate bilinear form $\left( \cdot ,\cdot
\right) _{\mathfrak{h}}:\mathfrak{h}\times \mathfrak{h}\rightarrow \mathbb{R}
$, see eq. $\left( \ref{Lie alg g+g 2}\right) $, has the following non
trivial pairings 
\begin{eqnarray*}
\left( \left( e_{2},0\right) ,\left( 0,e_{2}\right) \right) _{\mathfrak{h}}
&=&\left( \left( e_{4},0\right) ,\left( 0,e_{4}\right) \right) _{\mathfrak{h}%
}=1 \\
\left( \left( e_{1},0\right) ,\left( 0,e_{3}\right) \right) _{\mathfrak{h}}
&=&\left( \left( e_{3},0\right) ,\left( 0,e_{1}\right) \right) _{\mathfrak{h}%
}=-1
\end{eqnarray*}%
giving rise to the \emph{Manin triple} $\left( \mathfrak{h},\mathfrak{h}_{+},%
\mathfrak{h}_{-}\right) $ with 
\begin{eqnarray*}
\mathfrak{h}_{+} &=&\mathfrak{g}_{+}\oplus \mathfrak{g}_{+}^{\bot
}=lspan\left\{ \left( e_{2},0\right) ,\left( e_{3},0\right) ,\left(
e_{4},0\right) ,\left( 0,e_{3}\right) \right\} \\
\mathfrak{h}_{-} &=&\mathfrak{g}_{-}\oplus \mathfrak{g}_{-}^{\bot
}=lspan\left\{ \left( e_{1},0\right) ,\left( 0,e_{1}\right) ,\left(
0,e_{2}\right) ,\left( 0,e_{4}\right) \right\}
\end{eqnarray*}

The Lie subalgebras $\mathfrak{h}_{+},\mathfrak{h}_{-}$ are indeed \emph{Lie
bialgebras} with the non vanishing Lie brackets and cobrackets 
\begin{eqnarray}
\mathfrak{h}_{+} &\longrightarrow &\left\{ 
\begin{array}{l}
\left[ \left( e_{4},0\right) ,\left( e_{2},0\right) \right] =\left(
e_{3},0\right) \\ 
\delta _{+}\left( e_{4},0\right) =\left( e_{3},0\right) \wedge \left(
0,e_{3}\right) -\left( 0,e_{3}\right) \wedge \left( e_{2},0\right)%
\end{array}%
\right.  \notag \\
&&  \label{3step 4} \\
\mathfrak{h}_{-} &\longrightarrow &\left\{ 
\begin{array}{l}
\left[ \left( e_{1},0\right) ,\left( 0,e_{2}\right) \right] =\left(
0,e_{4}\right) \\ 
\delta _{-}\left( 0,e_{1}\right) =\left( 0,e_{2}\right) \wedge \left(
0,e_{4}\right)%
\end{array}%
\right.  \notag
\end{eqnarray}%
so the associated Lie groups $H_{\pm }=G_{\pm }\oplus \mathfrak{g}_{\pm
}^{\bot }$ become into \emph{Poisson Lie groups. }We determine the Poisson
Lie structure in the matrix representation of this Lie algebra $\left( \ref%
{3step 0}\right) $ in a four dimensional vector space where the Lie algebra
generators, in terms of the $4\times 4$ elementary matrices $\left(
E_{ij}\right) _{kl}=\delta _{ik}\delta _{jl}$, are

\begin{equation*}
\begin{array}{ccccccc}
e_{1}=E_{34} & , & e_{2}=E_{24} & , & e_{3}=E_{14} & , & e_{4}=E_{12}+E_{23}%
\end{array}%
\end{equation*}%
In this representation, any vector $X=\left( x_{1},x_{2},x_{3},x_{4}\right) $
of this Lie algebra is 4-step nilpotent, $X^{4}=0$.

The \emph{Poisson-Lie bivector} $\pi _{+}$ on $H_{+}$ is defined by the
relation $\left( \ref{PL bivector h+ 1}\right) $. Since the exponential map
is surjective on nilpotent Lie groups, we may write 
\begin{eqnarray*}
\left( g_{+},Z_{+}^{\bot }\right) &=&\left(
e^{u_{2}e_{2}+u_{3}e_{3}+u_{4}e_{4}},z_{3}e_{3}\right) \\
\left( X_{-},Y_{-}^{\bot }\right) &=&\left(
x_{1}e_{1},y_{1}e_{1}+y_{2}e_{2}+y_{4}e_{4}\right)
\end{eqnarray*}%
to get 
\begin{eqnarray*}
&&\left\langle \gamma \left( X_{-}^{\prime },Y_{-}^{\prime \bot }\right)
\left( g_{+},Z_{+}^{\bot }\right) ^{-1}\otimes \gamma \left( X_{-}^{\prime
\prime },Y_{-}^{\prime \prime \bot }\right) \left( g_{+},Z_{+}^{\bot
}\right) ^{-1},\pi _{+}\left( g_{+},Z_{+}^{\bot }\right) \right\rangle \\
&=&\frac{1}{2}y_{1}^{\prime }u_{4}^{2}x_{1}^{\prime \prime }-\frac{1}{2}%
x_{1}^{\prime }u_{4}^{2}y_{1}^{\prime \prime }+x_{1}^{\prime
}u_{4}y_{2}^{\prime \prime }-y_{2}^{\prime }u_{4}x_{1}^{\prime \prime }
\end{eqnarray*}%
Introducing $\pi _{+}^{R}:G\longrightarrow \mathfrak{h}_{+}\otimes \mathfrak{%
h}_{+}$ as $\pi _{+}^{R}\left( g_{+},Z_{+}^{\bot }\right) =\left( R_{\left(
g_{+},Z_{+}^{\bot }\right) ^{-1}}\right) _{\ast }^{\otimes 2}\pi _{+}\left(
g_{+},Z_{+}^{\bot }\right) $, we get 
\begin{eqnarray*}
\pi _{+}^{R}\left( g_{+},Z_{+}^{\bot }\right) &=&\frac{1}{2}u_{4}^{2}\left(
e_{3},0\right) \otimes \left( 0,e_{3}\right) -\frac{1}{2}u_{4}^{2}\left(
0,e_{3}\right) \otimes \left( e_{3},0\right) \\
&&-u_{4}\left( 0,e_{3}\right) \otimes \left( e_{2},0\right) +u_{4}\left(
e_{2},0\right) \otimes \left( 0,e_{3}\right)
\end{eqnarray*}%
The dressing vector can be obtained from the PL bivector by using eq. $%
\left( \ref{PL bivector h+ 7}\right) $ to get 
\begin{equation*}
\left( g_{+},Z_{+}^{\bot }\right) ^{\left( X_{-}^{\prime \prime
},Y_{-}^{\prime \prime \bot }\right) }=\left( -u_{4}x_{1}^{\prime \prime
}e_{2}-\frac{1}{2}u_{4}^{2}x_{1}^{\prime \prime }e_{3},\left( \frac{1}{2}%
u_{4}^{2}y_{1}^{\prime \prime }-u_{4}y_{2}^{\prime \prime }\right)
e_{3}\right)
\end{equation*}%
and, the cobracket $\delta _{+}:\mathfrak{h}_{+}\longrightarrow \mathfrak{h}%
_{+}\wedge \mathfrak{h}_{+}$ defined from $\left( \pi _{+}^{R}\right) _{\ast
\left( e,0\right) }$ reproducing $\left( \ref{3step 4}\right) $. Since $%
\mathfrak{h}_{+}$ has only one non trivial Lie bracket, see eq. $\left( \ref%
{3step 4}\right) $, there is no an object $r\in \mathfrak{h}_{+}\otimes 
\mathfrak{h}_{+}$ making $\delta _{+}$ in a coboundary.

In order to determine \emph{the Poisson-Lie structure on} $H_{-}$, we write
the elements in $H_{-}$ and $\mathfrak{h}_{+}$ as 
\begin{eqnarray*}
\left( g_{-},Z_{-}^{\bot }\right) &=&\left(
e^{u_{1}e_{1}},z_{1}e_{1}+z_{2}e_{2}+z_{4}e_{4}\right) \\
\left( X_{+},Y_{+}^{\bot }\right) &=&\left(
x_{2}e_{2}+x_{3}e_{3}+x_{4}e_{4},y_{3}e_{3}\right)
\end{eqnarray*}%
and PL bivector $\pi _{-}$ on $H_{-}$ defined by the relation $\left( \ref%
{PL bivector h- 1}\right) $ is 
\begin{equation*}
\left\langle \left( g_{-},Z_{-}^{\bot }\right) ^{-1}\left( \gamma \left(
X_{+}^{\prime },Y_{+}^{\prime \bot }\right) \otimes \gamma \left(
X_{+}^{\prime \prime },Y_{+}^{\prime \prime \bot }\right) \right) ,\pi
_{-}\left( g_{-},Z_{-}^{\bot }\right) \right\rangle =x_{4}^{\prime
}z_{1}x_{2}^{\prime \prime }-x_{2}^{\prime }z_{1}x_{4}^{\prime \prime }
\end{equation*}

Writing it in terms of the dressing vector as in eq. $\left( \ref{PL
bivector h- 2}\right) $ we obtain%
\begin{equation*}
\left( g_{-},Z_{-}^{\bot }\right) ^{\left( X_{+}^{\prime \prime
},Y_{+}^{\prime \prime \bot }\right) }=\left( 0,-z_{1}x_{4}^{\prime \prime
}e_{2}+z_{1}x_{2}^{\prime \prime }e_{4}\right)
\end{equation*}%
Introducing $\pi _{-}^{L}:G\longrightarrow \mathfrak{h}_{-}\otimes \mathfrak{%
h}_{-}$ as $\pi _{-}^{L}\left( g_{-},Z_{-}^{\bot }\right) =\left( L_{\left(
g_{-},Z_{-}^{\bot }\right) ^{-1}}\right) _{\ast }^{\otimes 2}\pi _{-}\left(
g_{-},Z_{-}^{\bot }\right) $, we have%
\begin{equation*}
\pi _{-}^{L}\left( g_{-},Z_{-}^{\bot }\right) =z_{1}\left( 0,e_{4}\right)
\otimes \left( 0,e_{2}\right) -z_{1}\left( 0,e_{2}\right) \otimes \left(
0,e_{4}\right)
\end{equation*}%
The cobracket on $\mathfrak{h}_{-}$, $\delta _{-}:\mathfrak{h}%
_{-}\longrightarrow \mathfrak{h}_{-}\wedge \mathfrak{h}_{-}$ given in $%
\left( \ref{3step 4}\right) $ can be easily retrieved as $\delta
_{-}:=-\left( \pi _{-}^{L}\right) _{\ast \left( e,0\right) }$. Again, the
only nonvanishing Lie bracket in $\mathfrak{h}_{-}$ does not allows for some
object $r$ in $\mathfrak{h}_{-}\otimes \mathfrak{h}_{-}$ giving rise to that
cobracket.

The decomposition $\mathfrak{h}=\mathfrak{h}_{+}\oplus \mathfrak{h}_{-}$
implies the factorization of the Lie group $H$ as $H_{+}H_{-}$ such that%
\begin{eqnarray*}
&&\bigskip \left(
e^{u_{1}e_{1}+u_{2}e_{2}+u_{3}e_{3}+u_{4}e_{4}},z_{1}e_{1}+z_{2}e_{2}+z_{3}e_{3}+z_{4}e_{4}\right)
\\
&=&\left( e^{\left( u_{2}-\frac{1}{2}u_{1}u_{4}\right) e_{2}+\left( u_{3}-%
\frac{1}{12}u_{1}u_{4}^{2}\right) e_{3}+u_{4}e_{4}},z_{3}e_{3}\right) \cdot
\left( e^{u_{1}e_{1}},z_{1}e_{1}+z_{2}e_{2}+z_{4}e_{4}\right)
\end{eqnarray*}%
and from here we get the reciprocal dressing actions%
\begin{equation*}
\left\{ 
\begin{array}{l}
\left( e^{u_{2}e_{2}+u_{3}e_{3}+u_{4}e_{4}},z_{3}e_{3}\right) ^{\left(
e^{u_{1}e_{1}},z_{1}e_{1}+z_{2}e_{2}+z_{4}e_{4}\right) } \\ 
=\left( e^{^{\left( u_{2}-u_{1}u_{4}\right) e_{2}+\left( u_{3}-\frac{1}{2}%
u_{1}u_{4}^{2}\right) e_{3}+u_{4}e_{4}}},\left( \dfrac{1}{2}%
z_{1}u_{4}^{2}-z_{2}u_{4}+z_{3}\right) e_{3}\right) \\ 
\\ 
\left( e^{u_{1}e_{1}},z_{1}e_{1}+z_{2}e_{2}+z_{4}e_{4}\right) ^{\left(
e^{u_{2}e_{2}+u_{3}e_{3}+u_{4}e_{4}},z_{3}e_{3}\right) } \\ 
=\left( e^{u_{1}e_{1}},z_{1}e_{1}+\left( z_{2}-z_{1}u_{4}\right)
e_{2}+\left( z_{1}u_{2}+z_{4}\right) e_{4}\right)%
\end{array}%
\right.
\end{equation*}

\subsubsection{AKS Integrable system}

In order to produces an AKS integrable systems, we consider an $\mathrm{ad}^{%
\mathfrak{h}}$-invariant function $\mathsf{H}$ on $\mathfrak{h}$. We
consider $\left( \cdot ,\cdot \right) _{\mathfrak{h}}:\mathfrak{h}\times 
\mathfrak{h}\rightarrow \mathbb{R}$%
\begin{equation*}
\mathsf{H}\left( X,Y\right) =\frac{1}{2}\left( \left( \left( X,Y\right)
,\left( X,Y\right) \right) _{\mathfrak{h}}\right) =\left( X,Y\right) _{%
\mathfrak{g}}
\end{equation*}%
so 
\begin{equation*}
\mathfrak{L}_{\mathsf{H}}\left( X,Y\right) =\left( X,Y\right)
\end{equation*}

The hamiltonian vector field is%
\begin{equation*}
V_{\mathsf{H}}\left( X,Y\right) =\left[ \left( X_{-},Y_{-}^{\bot }\right)
,\left( X_{+},Y_{+}^{\bot }\right) \right]
\end{equation*}%
The condition $\psi \left( X_{-},Y_{-}^{\bot }\right) \in \emph{char}\left( 
\mathfrak{h}_{+}\right) $, is satisfied provided the component in $\left(
0,e_{1}\right) $ vanish, so the allowed motion is performed on points 
\begin{equation*}
\left( X,Y\right) =\left(
x_{1}e_{1}+x_{2}e_{2}+x_{3}e_{3}+x_{4}e_{4},y_{2}e_{2}+y_{3}e_{3}+y_{4}e_{4}%
\right)
\end{equation*}%
and the Hamiltonian vector fields reduces to%
\begin{equation*}
V_{\mathsf{H}}\left( Z\right) =-x_{4}\left( x_{1}e_{2},y_{2}e_{3}\right)
\end{equation*}%
giving rise to non trivial Hamilton equation of motion $\dot{Z}\left(
t\right) =V_{\mathsf{H}}\left( Z\left( t\right) \right) $ 
\begin{equation}
\left\{ 
\begin{array}{l}
\dot{x}_{2}=-x_{4}x_{1}\medskip \\ 
\dot{y}_{3}=-x_{4}y_{2}%
\end{array}%
\right.  \label{3s07}
\end{equation}%
The coordinates $x_{1},x_{3},x_{4},y_{2},y_{4}$ remains constants of motion,
unveiling a quite simple dynamical system, namely a uniform linear motion in
the plane $\left( x_{2},y_{3}\right) $. Despite this simplicity, it is
interesting to see how the $H_{+}$ factor of the semidirect product
exponential curve succeeds in yielding this linear trajectories as the
orbits by the adjoint action.

In order to apply the AKS Theorem, we obtain%
\begin{equation*}
\mathrm{Exp}^{\cdot }\left( t\mathfrak{L}_{\mathsf{H}}(X_{+\circ
}+K_{-})\right) =h_{+}\left( t\right) h_{-}\left( t\right)
\end{equation*}%
with%
\begin{eqnarray*}
h_{+}\left( t\right) &=&\left( e^{\left( tx_{2}-\frac{1}{2}%
t^{2}x_{1}x_{4}\right) e_{2}+\left( tx_{3}-\frac{1}{12}t^{3}x_{1}x_{4}^{2}%
\right) e_{3}+tx_{4}e_{4}},-\left( \frac{1}{2}t^{2}x_{4}y_{2}-ty_{3}\right)
e_{3}\right) \\
h_{-}\left( t\right) &=&\left( e^{tx_{1}e_{1}},ty_{2}e_{2}-\left( \frac{1}{2}%
t^{2}x_{1}y_{2}-ty_{4}\right) e_{4}\right)
\end{eqnarray*}

Thus, for the initial condition%
\begin{equation*}
Z\left( t_{0}\right) =\left(
x_{10}e_{1}+x_{20}e_{2}+x_{30}e_{3}+x_{40}e_{4},y_{20}e_{2}+y_{30}e_{3}+y_{40}e_{4}\right)
\end{equation*}%
the system $\left( \ref{3s07}\right) $ has the solution\textit{\ }%
\begin{equation*}
Z\left( t\right) =\mathrm{Ad}_{h_{+}^{-1}\left( t\right) }^{H}Z\left(
t_{0}\right)
\end{equation*}%
described by the curve 
\begin{equation*}
h_{+}^{-1}\left( t\right) =\left( e^{-\left( tx_{20}-\frac{t^{2}}{2}%
x_{10}x_{40}\right) e_{2}-\left( tx_{30}-\frac{t^{3}}{12}x_{10}x_{40}^{2}%
\right) e_{3}-tx_{40}e_{4}},\left( \frac{t^{2}}{2}x_{40}y_{20}-ty_{30}%
\right) e_{3}\right)
\end{equation*}%
The explicit form of the adjoint curve is 
\begin{eqnarray*}
Z\left( t\right) &=&\left( x_{10}e_{1}+\left( x_{20}-tx_{10}x_{40}\right)
e_{2}+x_{30}e_{3}+x_{40}e_{4}\right. , \\
&&\left. y_{20}e_{2}+\left( y_{30}-tx_{40}y_{20}\right)
e_{3}+y_{40}e_{4}\right)
\end{eqnarray*}%
solving the system of Hamilton equations $\left( \ref{3s07}\right) $.

\subsection{The Lie Group $G=\mathrm{A}_{6.34}$}

This example was taken from reference \cite{ghanam}. Its Lie algebra $%
\mathfrak{g}=\mathfrak{a}_{6.34}$ is generated by the basis $\left\{
e_{1},e_{2},e_{3},e_{4},e_{5},e_{6}\right\} $ with the non vanishing Lie
brackets%
\begin{equation*}
\begin{array}{ccccccc}
\left[ e_{2},e_{3}\right] =e_{1} & , & \left[ e_{2},e_{6}\right] =e_{3} & ,
& \left[ e_{3},e_{6}\right] =-e_{2} & , & \left[ e_{4},e_{6}\right] =e_{5}%
\end{array}%
\end{equation*}%
The Lie algebra can be represented by $6\times 6$ matrices, and by using the
elementary matrices $\left( E_{ij}\right) _{kl}=\delta _{ik}\delta _{jl}$
they can be written as 
\begin{equation*}
\left\{ 
\begin{array}{lllll}
e_{1}=-2E_{36} & , & e_{2}=E_{35} & , & e_{3}=E_{34}-E_{56} \\ 
e_{4}=E_{12} & , & e_{5}=E_{16} & , & e_{6}=E_{26}+E_{54}-E_{45}%
\end{array}%
\right. ,
\end{equation*}%
and a typical element $g\left( z,x,y,p,q,\theta \right) $ of the\ associated
Lie group $G=\mathrm{A}_{6.34}$ is%
\begin{equation*}
g\left( z,x,y,p,q,\theta \right) =\left( 
\begin{array}{cccccc}
1 & p & 0 & 0 & 0 & q \\ 
0 & 1 & 0 & 0 & 0 & \theta \\ 
0 & 0 & 1 & x\sin \theta +y\cos \theta & x\cos \theta -y\sin \theta & z \\ 
0 & 0 & 0 & \cos \theta & -\sin \theta & x \\ 
0 & 0 & 0 & \sin \theta & \cos \theta & -y \\ 
0 & 0 & 0 & 0 & 0 & 1%
\end{array}%
\right)
\end{equation*}

The group $G$ can be factorize as $G=G_{+}G_{-}$ with $G_{+},G_{-}$ being
the Lie subgroups 
\begin{eqnarray*}
G_{+} &=&\left\{ g_{+}\left( x,y,z\right) =g\left( z,x,y,0,0,0\right)
/\left( x,y,z\right) \in \mathbb{R}^{3}\right\} \\
&& \\
G_{-} &=&\left\{ g_{-}\left( p,q,\theta \right) =g\left( 0,0,0,p,q,\theta
\right) /\left( p,q,\theta \right) \in \mathbb{R}^{3}\right\}
\end{eqnarray*}%
in such a way that, for $g\left( z,x,y,p,q,\theta \right) \in G$, 
\begin{equation*}
g\left( z,x,y,p,q,\theta \right) =g_{+}\left( x,y,z\right) g_{-}\left(
p,q,\theta \right) ~~.
\end{equation*}

By using this result we determine the dressing actions%
\begin{equation*}
\left\{ 
\begin{array}{l}
\left( g_{+}\left( x,y,z\right) \right) ^{g_{-}\left( p,q,\theta \right)
}=g_{+}\left( x\cos \theta +y\sin \theta ,y\cos \theta -x\sin \theta
,z\right) \\ 
\\ 
\left( g_{-}\left( p,q,\theta \right) \right) ^{g_{+}\left( x,y,z\right)
}=g_{-}\left( p,q,\theta \right)%
\end{array}%
\right. .
\end{equation*}

The Lie algebra $\mathfrak{g}$ decompose then as $\mathfrak{g}=\mathfrak{g}%
_{+}\oplus \mathfrak{g}_{-}$ with%
\begin{equation*}
\begin{array}{ccc}
\mathfrak{g}_{+}=Lspan\left\{ e_{1},e_{2},e_{3}\right\} & , & \mathfrak{g}%
_{-}=Lspan\left\{ e_{4},e_{5},e_{6}\right\}%
\end{array}%
\end{equation*}%
being Lie subalgebras of $\mathfrak{g}$. The exponential map is surjective
and each element $g\left( z,x,y,p,q,\theta \right) \in \mathrm{A}_{6.34}$
can be written as%
\begin{eqnarray*}
&&g\left( z,x,y,p,q,\theta \right) \\
&=&e^{\left( -\frac{1}{2}ze_{1}+\frac{\theta }{2}\frac{\left( x\sin \theta
-y\left( 1-\cos \theta \right) \right) }{\left( 1-\cos \theta \right) }e_{2}+%
\frac{\theta }{2}\frac{\left( x\left( 1-\cos \theta \right) +y\sin \theta
\right) }{\left( 1-\cos \theta \right) }e_{3}+pe_{4}+\left( q-\frac{1}{2}%
p\theta \right) e_{5}+\theta e_{6}\right) }
\end{eqnarray*}

Let us now introduce a bilinear form on $\mathfrak{g}$ inherited from the
non-invariant metric on $G$ given in reference \cite{ghanam}, 
\begin{equation*}
g=dp^{2}+\left( dq-\frac{1}{2}\left( pd\theta +\theta dp\right) \right)
^{2}+dx^{2}+dy^{2}-ydxd\theta +xdyd\theta +dzd\theta
\end{equation*}%
By taking this metric at the identity element of $G$ we get the
nondegenerate symmetric bilinear form $\left( ,\right) _{\mathfrak{g}}:%
\mathfrak{g}\otimes \mathfrak{g}\longrightarrow \mathbb{R}$ defined as 
\begin{eqnarray*}
&&\left( X\left( x_{1},x_{2},x_{3},x_{4},x_{5},x_{6}\right) ,X\left(
x_{1}^{\prime },x_{2}^{\prime },x_{3}^{\prime },x_{4}^{\prime
},x_{5}^{\prime },x_{6}^{\prime }\right) \right) _{\mathfrak{g}} \\
&=&x_{2}x_{2}^{\prime }+x_{3}x_{3}^{\prime }+x_{4}x_{4}^{\prime
}+x_{5}x_{5}^{\prime }+x_{1}x_{6}^{\prime }+x_{6}x_{1}^{\prime }
\end{eqnarray*}%
It gives rise to the decomposition of $\mathfrak{g}$ as $\mathfrak{g}%
_{+}^{\bot }\oplus \mathfrak{g}_{-}^{\bot }$ where 
\begin{equation*}
\begin{array}{ccc}
\mathfrak{g}_{+}^{\bot }=Lspan\left\{ e_{1},e_{4},e_{5}\right\} & , & 
\mathfrak{g}_{-}^{\bot }=Lspan\left\{ e_{2},e_{3},e_{6}\right\}%
\end{array}%
\end{equation*}%
This bilinear form defines associated $\tau $-action of $\mathfrak{g}$ on
itself 
\begin{equation*}
\begin{array}{lllllll}
\mathrm{ad}_{e_{3}}^{\tau }e_{2}=e_{1} & , & \mathrm{ad}_{e_{2}}^{\tau
}e_{3}=-e_{1} & , & \mathrm{ad}_{e_{4}}^{\tau }e_{5}=-e_{1} & , & \mathrm{ad}%
_{e_{2}}^{\tau }e_{6}=-e_{3} \\ 
\mathrm{ad}_{e_{6}}^{\tau }e_{2}=-e_{3} & , & \mathrm{ad}_{e_{6}}^{\tau
}e_{3}=e_{2} & , & \mathrm{ad}_{e_{6}}^{\tau }e_{5}=e_{4} & , & \mathrm{ad}%
_{e_{3}}^{\tau }e_{6}=e_{2}%
\end{array}%
\end{equation*}

So, let us now consider the semidirect sum $\mathfrak{h}=\mathfrak{g}\ltimes 
\mathfrak{g}$ where the left $\mathfrak{g}$ acts on the other one by the $%
\tau $-action. We have then the decomposition $\mathfrak{h}=\mathfrak{h}%
_{+}\oplus \mathfrak{h}_{-}$ with%
\begin{eqnarray*}
\mathfrak{h}_{+} &=&\mathfrak{g}_{+}\oplus \mathfrak{g}_{+}^{\bot
}=Lspan\left\{ \left( e_{1},0\right) ,\left( e_{2},0\right) ,\left(
e_{3},0\right) ,\left( 0,e_{1}\right) ,\left( 0,e_{4}\right) \left(
0,e_{5}\right) \right\} \\
&& \\
\mathfrak{h}_{-} &=&\mathfrak{g}_{-}\oplus \mathfrak{g}_{-}^{\bot
}=Lspan\left\{ \left( e_{4},0\right) ,\left( e_{5},0\right) ,\left(
e_{6},0\right) ,\left( 0,e_{2}\right) ,\left( 0,e_{3}\right) \left(
0,e_{6}\right) \right\}
\end{eqnarray*}%
The Lie algebra structure on $\mathfrak{h}$ is given by $\left( \ref{Lie alg
g+g 1}\right) $, with the fundamental Lie brackets%
\begin{equation*}
\begin{array}{lll}
\left[ \left( e_{2},0\right) ,\left( e_{3},0\right) \right] =\left(
e_{1},0\right) &  & \left[ \left( e_{3},0\right) ,\left( 0,e_{2}\right) %
\right] =\left( 0,e_{1}\right) \\ 
\left[ \left( e_{2},0\right) ,\left( e_{6},0\right) \right] =\left(
e_{3},0\right) &  & \left[ \left( e_{3},0\right) ,\left( 0,e_{6}\right) %
\right] =\left( 0,e_{2}\right) \\ 
\left[ \left( e_{3},0\right) ,\left( e_{6},0\right) \right] =-\left(
e_{2},0\right) &  & \left[ \left( e_{4},0\right) ,\left( 0,e_{5}\right) %
\right] =-\left( 0,e_{1}\right) \\ 
\left[ \left( e_{4},0\right) ,\left( e_{6},0\right) \right] =\left(
e_{5},0\right) &  & \left[ \left( e_{6},0\right) ,\left( 0,e_{2}\right) %
\right] =-\left( 0,e_{3}\right) \\ 
\left[ \left( e_{2},0\right) ,\left( 0,e_{3}\right) \right] =-\left(
0,e_{1}\right) &  & \left[ \left( e_{6},0\right) ,\left( 0,e_{3}\right) %
\right] =\left( 0,e_{2}\right) \\ 
\left[ \left( e_{2},0\right) ,\left( 0,e_{6}\right) \right] =-\left(
0,e_{3}\right) &  & \left[ \left( e_{6},0\right) ,\left( 0,e_{5}\right) %
\right] =\left( 0,e_{4}\right)%
\end{array}%
\end{equation*}%
The semidirect product structure of $H=G\ltimes \mathfrak{g}$ is defined by
eq. $\left( \ref{semidirect prod}\right) $.

Every element $\left( g\left( z,x,y,p,q,\theta \right)
,\sum_{i=1}^{6}x_{i}e_{i}\right) \in H$ can be factorized as 
\begin{eqnarray*}
&&\left( g\left( z,x,y,p,q,\theta \right) ,\sum_{i=1}^{6}x_{i}e_{i}\right) \\
&=&\left( g_{+}\left( z,x,y\right) ,\left( x_{1}-x_{5}p\right) e_{1}+\left(
x_{4}+x_{5}\theta \right) e_{4}+x_{5}e_{5}\right) \\
&&\cdot \left( g_{-}\left( p,q,\theta \right)
,x_{2}e_{2}+x_{3}e_{3}+x_{6}e_{6}\right)
\end{eqnarray*}%
from where we get the reciprocal \emph{dressing actions} 
\begin{equation*}
\begin{array}{l}
\left( g_{+}\left( x,y,z\right) ,x_{1}e_{1}+x_{4}e_{4}+x_{5}e_{5}\right)
^{\left( g_{-}\left( p,q,\theta \right)
,x_{2}e_{2}+x_{3}e_{3}+x_{6}e_{6}\right) } \\ 
~=\left( g_{+}\left( x\cos \theta +y\sin \theta ,y\cos \theta -x\sin \theta
,z\right) \right. , \\ 
~~~~~\left. \left( \frac{1}{2}x_{6}\left( x^{2}+y^{2}\right) -\left(
yx_{2}-xx_{3}\right) +x_{1}-x_{5}p\right) e_{1}+\left( x_{4}+x_{5}\theta
\right) e_{4}+x_{5}e_{5}\right)%
\end{array}%
\end{equation*}%
and%
\begin{equation*}
\begin{array}{l}
\left( g_{-}\left( p,q,\theta \right)
,x_{2}e_{2}+x_{3}e_{3}+x_{6}e_{6}\right) ^{\left( g_{+}\left( x,y,z\right)
,x_{1}e_{1}+x_{4}e_{4}+x_{5}e_{5}\right) } \\ 
~=\left( g_{-}\left( p,q,\theta \right) ,\left( x_{2}-x_{6}y\right)
e_{2}+\left( x_{3}+x_{6}x\right) e_{3}+x_{6}e_{6}\right)%
\end{array}%
\end{equation*}%
The infinitesimal generator of these dressing action are 
\begin{equation*}
\begin{array}{l}
\left( g_{+}\left( x,y,z\right) ,x_{1}e_{1}+x_{4}e_{4}+x_{5}e_{5}\right)
^{\left( z_{4}e_{4}+z_{5}e_{5}+z_{6}e_{6},z_{2}^{\prime }e_{2}+z_{3}^{\prime
}e_{3}+z_{6}^{\prime }e_{6}\right) } \\ 
~=\left( yz_{6}e_{2}-xz_{6}e_{3},\left( \frac{1}{2}\left( x^{2}+y^{2}\right)
z_{6}^{\prime }-x_{5}z_{4}+xz_{3}^{\prime }-yz_{2}^{\prime }\right)
e_{1}+x_{5}z_{6}e_{4}\right)%
\end{array}%
\end{equation*}%
and

\begin{equation*}
\begin{array}{l}
\left( g_{-}\left( p,q,\theta \right)
,x_{2}e_{2}+x_{3}e_{3}+x_{6}e_{6}\right) ^{\left(
z_{1}e_{1}+z_{2}e_{2}+z_{3}e_{3},z_{1}^{\prime }e_{1}+z_{4}^{\prime
}e_{4}+z_{5}^{\prime }e_{5}\right) } \\ 
~=\left( 0,-x_{6}z_{3}e_{2}+x_{6}z_{2}e_{3}\right)%
\end{array}%
\end{equation*}

The Poisson-Lie bivector is determined from the relation $\left( \ref{PL
bivector h+ 1}\right) $, that for 
\begin{eqnarray*}
\left( g_{+},X_{+}^{\bot }\right) &=&\left( g\left( z,x,y\right)
,x_{1}e_{1}+x_{4}e_{4}+x_{5}e_{5}\right) \\
\left( Y_{-},Y_{-}^{\bot }\right) &=&\left(
y_{4}e_{4}+y_{5}e_{5}+y_{6}e_{6},y_{2}^{\prime }e_{2}+y_{3}^{\prime
}e_{3}+y_{6}^{\prime }e_{6}\right) \\
\left( Z_{-},Z_{-}^{\bot }\right) &=&\left(
z_{4}e_{4}+z_{5}e_{5}+z_{6}e_{6},z_{2}^{\prime }e_{2}+z_{3}^{\prime
}e_{3}+z_{6}^{\prime }e_{6}\right)
\end{eqnarray*}%
we get 
\begin{eqnarray*}
&&\left\langle \gamma \left( Y_{-},Y_{-}^{\bot }\right) \left(
g_{+},Z_{+}^{\bot }\right) ^{-1}\otimes \gamma \left( Z_{-},Z_{-}^{\bot
}\right) \left( g_{+},X_{+}^{\bot }\right) ^{-1},\pi _{+}\left(
g_{+},X_{+}^{\bot }\right) \right\rangle \\
&=&y_{4}x_{5}z_{6}+y_{6}xz_{3}^{\prime }-y_{6}x_{5}z_{4}-y_{6}yz_{2}^{\prime
} \\
&&+\frac{1}{2}y_{6}\left( x^{2}+y^{2}\right) z_{6}^{\prime }+y_{2}^{\prime
}yz_{6}-y_{3}^{\prime }xz_{6}-\frac{1}{2}y_{6}^{\prime }\left(
x^{2}+y^{2}\right) z_{6}
\end{eqnarray*}%
Introducing the linear map $\pi _{+}^{R}:\mathfrak{h}_{-}\longrightarrow 
\mathfrak{h}_{+}$ defined as 
\begin{eqnarray*}
&&\left\langle \gamma \left( Y_{-},Y_{-}^{\bot }\right) \left(
g_{+},Z_{+}^{\bot }\right) ^{-1}\otimes \gamma \left( Z_{-},Z_{-}^{\bot
}\right) \left( g_{+},X_{+}^{\bot }\right) ^{-1},\pi _{+}\left(
g_{+},X_{+}^{\bot }\right) \right\rangle \\
&=&\left( \left( Y_{-},Y_{-}^{\bot }\right) ,\pi _{+}^{R}\left(
Z_{-},Z_{-}^{\bot }\right) \right) _{\mathfrak{h}}
\end{eqnarray*}%
we get that it is characterized by the matrix 
\begin{eqnarray*}
&&\pi _{+}^{R}\left( g\left( z,x,y\right)
,x_{1}e_{1}+x_{4}e_{4}+x_{5}e_{5}\right) \\
&& \\
&=&\left( 
\begin{array}{cccccc}
0 & 0 & x_{5} & 0 & 0 & 0 \\ 
0 & 0 & 0 & 0 & 0 & 0 \\ 
-x_{5} & 0 & 0 & -y & x & \frac{1}{2}\left( x^{2}+y^{2}\right) \\ 
0 & 0 & y & 0 & 0 & 0 \\ 
0 & 0 & -x & 0 & 0 & 0 \\ 
0 & 0 & -\frac{1}{2}\left( x^{2}+y^{2}\right) & 0 & 0 & 0%
\end{array}%
\right)
\end{eqnarray*}

The Poisson-Lie bivector $H_{-}$ is determined from the relation $\left( \ref%
{PL bivector h- 1}\right) $ for 
\begin{eqnarray*}
\left( g_{-},X_{-}^{\bot }\right) &=&\left( g\left( p,q,\theta \right)
,x_{2}e_{2}+x_{3}e_{3}+x_{6}e_{6}\right) \\
\left( Y_{+},Y_{+}^{\bot }\right) &=&\left(
y_{1}e_{1}+y_{2}e_{2}+y_{3}e_{3},y_{1}^{\prime }e_{1}+y_{4}^{\prime
}e_{4}+y_{5}^{\prime }e_{5}\right) \\
\left( Z_{+},Z_{+}^{\bot }\right) &=&\left(
z_{1}e_{1}+z_{2}e_{2}+z_{3}e_{3},z_{1}^{\prime }e_{1}+z_{4}^{\prime
}e_{4}+z_{5}^{\prime }e_{5}\right)
\end{eqnarray*}%
and after some lengthy computations we get 
\begin{equation*}
\left\langle \left( g_{-},X_{-}^{\bot }\right) ^{-1}\left( \gamma \left(
Y_{+},Y_{+}^{\bot }\right) \otimes \gamma \left( Z_{+},Z_{+}^{\bot }\right)
\right) ,\pi _{-}\left( g_{-},Z_{-}^{\bot }\right) \right\rangle
=y_{3}x_{6}z_{2}-y_{2}x_{6}z_{3}
\end{equation*}%
It has associated a bilinear form on $\mathfrak{h}_{+}$, that in the given
basis is characterized by the matrix 
\begin{equation*}
\pi _{-}^{L}\left( g\left( p,q,\theta \right)
,x_{2}e_{2}+x_{3}e_{3}+x_{6}e_{6}\right) =\left( 
\begin{array}{cccccc}
0 & 0 & 0 & 0 & 0 & 0 \\ 
0 & 0 & -x_{6} & 0 & 0 & 0 \\ 
0 & x_{6} & 0 & 0 & 0 & 0 \\ 
0 & 0 & 0 & 0 & 0 & 0 \\ 
0 & 0 & 0 & 0 & 0 & 0 \\ 
0 & 0 & 0 & 0 & 0 & 0%
\end{array}%
\right)
\end{equation*}

\subsubsection{AKS-Integrable system on $G=\mathrm{A}_{6.34}$}

Let us study a dynamical system on $\mathfrak{h}$ ruled by the $\mathrm{ad}^{%
\mathfrak{h}}$-invariant Hamilton function $\mathsf{H}$ on $\mathfrak{h}$%
\begin{equation*}
\mathsf{H}\left( X,X^{\prime }\right) =\frac{1}{2}\left( \left( \left(
X,X^{\prime }\right) ,\left( X,X^{\prime }\right) \right) _{\mathfrak{h}%
}\right) =\left( X,X^{\prime }\right) _{\mathfrak{g}}
\end{equation*}%
Writing $\left( X,X^{\prime }\right) \in \mathfrak{h}$ as 
\begin{equation*}
\left( X,X^{\prime }\right) =\left(
\sum_{i=1}^{6}x_{i}e_{i},\sum_{i=1}^{6}x_{i}^{\prime }e_{i}\right)
\end{equation*}%
it becomes in 
\begin{equation*}
\mathsf{H}\left( X,X^{\prime }\right) =\frac{1}{n}\left( x_{2}^{\prime
}x_{2}+x_{3}^{\prime }x_{3}+x_{4}^{\prime }x_{4}+x_{5}^{\prime
}x_{5}+x_{1}x_{6}^{\prime }+x_{6}x_{1}^{\prime }\right)
\end{equation*}%
for which 
\begin{equation*}
\mathfrak{L}_{\mathsf{H}}\left( X,X^{\prime }\right) =\left( X,X^{\prime
}\right)
\end{equation*}

The hamiltonian vector field is then%
\begin{equation*}
V_{\mathsf{H}}\left( X,X^{\prime }\right) =-\mathrm{ad}_{\Pi _{+}\mathfrak{L}%
_{\mathsf{H}}\left( X,X^{\prime }\right) }^{\mathfrak{h}}\left( X,X^{\prime
}\right) =\left[ \Pi _{-}\left( X,X^{\prime }\right) ,\Pi _{+}\left(
X,X^{\prime }\right) \right]
\end{equation*}%
Then, the explicit computation of the Lie bracket and applying the condition 
$\gamma \left( K_{-}\right) \in \emph{char}\left( \mathfrak{h}_{+}\right) $,
which means that $x_{6}^{\prime }=0$, we get the Hamiltonian vector fields 
\begin{equation*}
V_{\mathsf{H}}\left( X,X^{\prime }\right) =\left(
x_{3}x_{6}e_{2}-x_{2}x_{6}e_{3},\left( x_{2}x_{3}^{\prime
}-x_{3}x_{2}^{\prime }-x_{4}x_{5}^{\prime }\right) e_{1}+x_{5}^{\prime
}x_{6}e_{4}\right)
\end{equation*}%
and the nontrivial equation of motions are then%
\begin{equation}
\left\{ 
\begin{array}{l}
\dot{x}_{2}=x_{3}x_{6} \\ 
\dot{x}_{3}=-x_{2}x_{6} \\ 
\dot{x}_{1}^{\prime }=\left( x_{2}x_{3}^{\prime }-x_{3}x_{2}^{\prime
}-x_{4}x_{5}^{\prime }\right) \\ 
\dot{x}_{4}^{\prime }=x_{5}^{\prime }x_{6}%
\end{array}%
\right.  \label{A6.34 ham eqs}
\end{equation}%
with $x_{1},x_{4},x_{5},x_{6},x_{2}^{\prime },x_{3}^{\prime },x_{5}^{\prime
},x_{6}^{\prime }$ being time independent.

Let us now to integrate these equations by the AKS method. In doing so, we
need to factorize the exponential curve $\left( \ref{exp semid transad}%
\right) $%
\begin{equation*}
\mathrm{Exp}^{\cdot }t\mathfrak{L}_{\mathsf{H}}\left( X_{0},X_{0}^{\prime
}\right) =\mathrm{Exp}^{\cdot }t\left(
\sum_{i=1}^{6}x_{i0}e_{i},\sum_{i=1}^{5}x_{i0}^{\prime }e_{i}\right)
\end{equation*}%
for some initial $\left( X_{0},X_{0}^{\prime }\right) \in \mathfrak{h}$.

Factorizing the exponential 
\begin{equation*}
\mathrm{Exp}^{\cdot }t\mathfrak{L}_{\mathsf{H}}\left( X_{0},X_{0}^{\prime
}\right) =h_{+}\left( t\right) h_{-}\left( t\right)
\end{equation*}%
we obtain the curve%
\begin{equation*}
t\longmapsto h_{+}\left( t\right) =\left( g_{+}\left( t\right) ,W_{+}^{\bot
}\left( t\right) \right)
\end{equation*}%
where%
\begin{equation*}
g_{+}\left( t\right) =g_{+}\left( -2tx_{10},\tfrac{\left( x_{30}\left(
1-\cos tx_{60}\right) +x_{20}\sin tx_{60}\right) }{x_{60}},\tfrac{\left(
x_{20}\left( \cos tx_{60}-1\right) +x_{30}\sin tx_{60}\right) }{x_{60}}%
\right)
\end{equation*}%
and%
\begin{equation*}
W_{+}^{\bot }\left( t\right) =\left( tx_{10}^{\prime
}-2t^{2}x_{40}x_{50}^{\prime }\right) e_{1}+\left( tx_{4}^{\prime
}+2t^{2}x_{60}x_{50}^{\prime }\right) e_{4}+tx_{5}^{\prime }e_{5}
\end{equation*}%
which through the adjoint action%
\begin{equation*}
Z\left( t\right) =\mathrm{Ad}_{h_{+}^{-1}\left( t\right) }^{H}\left(
X_{0},X_{0}^{\prime }\right) =\left( \mathrm{Ad}_{g_{+}^{-1}\left( t\right)
}^{G}X_{0}\,,\tau _{g_{+}^{-1}\left( t\right) }\left( X_{0}^{\prime }-%
\mathrm{ad}_{X_{0}}^{\tau }W_{+}^{\bot }\left( t\right) \right) \right)
\end{equation*}%
gives rise to the solution of the Hamilton equations $\left( \ref{A6.34 ham
eqs}\right) $. In fact, after some calculations we get the nontrivial
solutions 
\begin{equation*}
\left\{ 
\begin{array}{l}
x_{2}\left( t\right) =x_{20}\cos tx_{60}+x_{30}\sin tx_{60} \\ 
x_{3}\left( t\right) =x_{30}\cos tx_{60}-x_{20}\sin tx_{60} \\ 
x_{1}^{\prime }\left( t\right) =\left( x_{10}^{\prime
}-tx_{40}x_{50}^{\prime }\right) -\dfrac{\left( x_{20}^{\prime }x_{2}\left(
t\right) +x_{30}^{\prime }x_{3}\left( t\right) \right) }{x_{60}}+\dfrac{%
\left( x_{30}x_{30}^{\prime }+x_{20}x_{20}^{\prime }\right) }{x_{60}} \\ 
x_{4}^{\prime }\left( t\right) =x_{40}^{\prime }+tx_{60}x_{50}^{\prime }%
\end{array}%
\right.
\end{equation*}

\subsection{$\mathfrak{sl}_{2}\left( \mathbb{C}\right) $ equipped with an
inner product}

Let us consider the Lie algebra $\mathfrak{sl}_{2}(\mathbb{C})$, with the
associated decomposition 
\begin{equation}
\mathfrak{sl}_{2}(\mathbb{C})^{\mathbb{R}}=\mathfrak{su}_{2}\oplus \mathfrak{%
b}  \label{iwasawa}
\end{equation}%
where $\mathfrak{b}$ is the subalgebra of upper triangular matrices with
real diagonal and null trace, and $\mathfrak{su}_{2}$ is the real subalgebra
of $\mathfrak{sl}_{2}(\mathbb{C})$ of antihermitean matrices. For $\mathfrak{%
su}_{2}$ we take the basis 
\begin{equation*}
X_{1}=\left( 
\begin{array}{cc}
0 & i \\ 
i & 0%
\end{array}%
\right) \quad ,\quad X_{2}=\left( 
\begin{array}{cc}
0 & 1 \\ 
-1 & 0%
\end{array}%
\right) \quad ,\quad X_{3}=\left( 
\begin{array}{cc}
i & 0 \\ 
0 & -i%
\end{array}%
\right)
\end{equation*}%
and in $\mathfrak{b}$ this one%
\begin{equation*}
E=\left( 
\begin{array}{cc}
0 & 1 \\ 
0 & 0%
\end{array}%
\right) \quad ,\quad iE=\left( 
\begin{array}{cc}
0 & i \\ 
0 & 0%
\end{array}%
\right) \quad ,\quad H=\left( 
\begin{array}{cc}
1 & 0 \\ 
0 & -1%
\end{array}%
\right)
\end{equation*}

The Killing form in $\mathfrak{sl}_{2}(\mathbb{C})$ is%
\begin{equation*}
\kappa (X,Y)=\mathrm{tr}{(ad}\left( {X}\right) {ad}\left( {Y}\right) {)}=4%
\mathrm{tr}{(XY)},
\end{equation*}%
and the bilinear form on $\mathfrak{sl}_{2}(\mathbb{C})$ 
\begin{equation*}
\mathrm{k}_{0}(X,Y)=-\frac{1}{4}\mathrm{\mathrm{Im}} \kappa (X,Y)
\end{equation*}%
is nondegenerate, symmetric and Ad-invariant, turning $\mathfrak{b}$\emph{\ }%
and\emph{\ }$\mathfrak{su}_{2}$\emph{\ into isotropic subspaces}. However,
it fails in to be an \emph{inner product} and, in consequence, it does not
give rise to a Riemannian metric on the Lie group $SL\left( 2,\mathbb{C}%
\right) $.

By introducing the idempotent linear operator $\mathcal{E}:\mathfrak{sl}_{2}(%
\mathbb{C})\longrightarrow \mathfrak{sl}_{2}(\mathbb{C})$ defined as 
\begin{equation*}
\begin{array}{lllll}
\mathcal{E}X_{1}=-E & , & \mathcal{E}X_{2}=iE & , & \mathcal{E}X_{3}=-H \\ 
\mathcal{E}E=-X_{1} & , & \mathcal{E}iE=X_{2} & , & \mathcal{E}H=-X_{3}%
\end{array}%
\end{equation*}%
we define the \emph{non Ad-invariant} \emph{inner product} $\mathrm{g}:%
\mathfrak{sl}_{2}(\mathbb{C})\otimes \mathfrak{sl}_{2}(\mathbb{C}%
)\longrightarrow \mathbb{R\ }$such that for $V,W\in \mathfrak{sl}_{2}(%
\mathbb{C})$ we have%
\begin{equation*}
\mathrm{g}\left( V,W\right) :=\mathrm{k}_{0}(V,\mathcal{E}W)~.
\end{equation*}%
So, by left or right translations, it induces a non bi-invariant Riemannian
metric on $SL\left( 2,\mathbb{C}\right) $, regarded as a real manifold.

Then, beside Iwasawa decomposition $\left( \ref{iwasawa}\right) $, we have
the vector subspace direct sum decomposition $\mathfrak{sl}_{2}(\mathbb{C})^{%
\mathbb{R}}=\left( \mathfrak{su}_{2}\right) ^{\bot }\oplus \mathfrak{b}%
^{\bot }$, where $^{\bot }$ stands for the orthogonal subspaces referred to
the inner product $\mathrm{g}$. Since the basis $\left\{
X_{1},X_{2},X_{3},E,\left( iE\right) ,H\right\} $ is an orthogonal one in
relation with $\mathrm{g}$, we have that $\left( \mathfrak{su}_{2}\right)
^{\bot }=\mathfrak{b}$ and $\mathfrak{b}^{\bot }=\mathfrak{su}_{2}$.

We now built an example of the construction developed in previous sections
by considering $\mathfrak{sl}_{2}(\mathbb{C})$ equipped with the inner
product $\mathrm{g}:\mathfrak{sl}_{2}(\mathbb{C})\otimes \mathfrak{sl}_{2}(%
\mathbb{C})\longrightarrow \mathbb{R}$. The $\tau $-action of $SL\left( 2,%
\mathbb{C}\right) $ on $\mathfrak{sl}_{2}(\mathbb{C})$ and of $\mathfrak{sl}%
_{2}(\mathbb{C})$ on itself are then 
\begin{equation*}
\begin{array}{ccc}
\tau _{g}=\mathcal{E}\circ \mathrm{Ad}_{g}\circ \mathcal{E} & , & \mathrm{ad}%
_{X}^{\tau }=\mathcal{E}\circ \mathrm{ad}_{X}\circ \mathcal{E}%
\end{array}%
\end{equation*}

In the remaining of this example we often denote $\mathfrak{g}_{+}=\mathfrak{%
su}_{2}$, $\mathfrak{g}_{-}=\mathfrak{b}$, $G_{+}=SU\left( 2\right) $ and $%
G_{-}=B$. Let us denote the elements $g\in SL(2,\mathbb{C)}$, $g_{+}\in
SU\left( 2\right) $ and $g_{-}\in B$ as 
\begin{equation*}
\begin{array}{ccccc}
g=\left( 
\begin{array}{cc}
\mu & \nu \\ 
\rho & \sigma%
\end{array}%
\right) & , & g_{+}=\left( 
\begin{array}{cc}
\alpha & \beta \\ 
-\bar{\beta} & \bar{\alpha}%
\end{array}%
\right) & , & g_{-}=\left( 
\begin{array}{cc}
a & z \\ 
0 & a^{-1}%
\end{array}%
\right)%
\end{array}%
,
\end{equation*}%
with $\mu ,\sigma ,\nu ,\rho ,\alpha ,\beta ,z\in \mathbb{C}$, $a\in \mathbb{%
R}_{>0}$, and $\mu \sigma -\nu \rho =\left\vert \alpha \right\vert
^{2}+\left\vert \beta \right\vert ^{2}=1$.

The factorization $SL(2,\mathbb{C)}=SU\left( 2\right) \times B$ means that $%
g=g_{+}g_{-}$ with 
\begin{equation*}
\begin{array}{ccc}
g_{+}=\frac{1}{\sqrt{\left\vert \mu \right\vert ^{2}+\left\vert \rho
\right\vert ^{2}}}\left( 
\begin{array}{cc}
\mu & -\bar{\rho} \\ 
\rho & \bar{\mu}%
\end{array}%
\right) & , & g_{-}\frac{1}{\sqrt{\left\vert \mu \right\vert ^{2}+\left\vert
\rho \right\vert ^{2}}}\left( 
\begin{array}{cc}
\left\vert \mu \right\vert ^{2}+\left\vert \rho \right\vert ^{2} & \bar{\mu}%
\nu +\bar{\rho}\sigma \\ 
0 & 1%
\end{array}%
\right)%
\end{array}%
\end{equation*}%
from where we determine the dressing actions 
\begin{equation*}
g_{+}^{g_{-}}=\frac{1}{\Delta \left( g_{+},g_{-}\right) }\left( 
\begin{array}{cc}
a\left( \alpha -\dfrac{1}{a}z\bar{\beta}\right) & \dfrac{1}{a}\beta \\ 
-\dfrac{1}{a}\bar{\beta} & a\left( \bar{\alpha}-\dfrac{1}{a}\bar{z}\beta
\right)%
\end{array}%
\right)
\end{equation*}%
\begin{equation*}
g_{-}^{g_{+}}=\left( 
\begin{array}{cc}
\Delta \left( g_{+},g_{-}\right) & \frac{\left( a^{2}\beta \bar{\alpha}+az%
\bar{\alpha}^{2}-a\bar{z}\beta ^{2}-\left( \left\vert z\right\vert ^{2}+%
\dfrac{1}{a^{2}}\right) \bar{\alpha}\beta \right) }{\Delta \left(
g_{+},g_{-}\right) } \\ 
0 & \dfrac{1}{\Delta \left( g_{+},g_{-}\right) }%
\end{array}%
\right)
\end{equation*}%
with 
\begin{equation*}
\Delta \left( g_{+},g_{-}\right) =\sqrt{a^{2}\left\vert \alpha \right\vert
^{2}-a\left( \bar{z}\alpha \beta +z\bar{\beta}\bar{\alpha}\right) +\left(
\left\vert z\right\vert ^{2}+\dfrac{1}{a^{2}}\right) \left\vert \beta
\right\vert ^{2}}~.
\end{equation*}

The dressing vector fields associated with $%
X_{+}=x_{1}X_{1}+x_{2}X_{2}+x_{3}X_{3}\in \mathfrak{g}_{+}$ is

\begin{equation*}
g_{-}^{X_{+}}=\dfrac{\sqrt{\mathrm{det}X_{+}}}{a}\left( 
\begin{array}{cc}
-a\left( bx_{2}+cx_{1}\right) & \Omega \left( x_{1},x_{2},x_{3},a,b,c\right)
\\ 
0 & \dfrac{1}{a}\left( bx_{2}+cx_{1}\right)%
\end{array}%
\right)
\end{equation*}%
where 
\begin{eqnarray*}
\Omega \left( x_{1},x_{2},x_{3},a,b,c\right)
&=&b^{2}x_{2}+ic^{2}x_{1}+2x_{3}a\left( c-ib\right) \\
&&+\left( bc-b^{2}-c^{2}-\dfrac{1}{a^{2}}\right) \left( ix_{1}+x_{2}\right)
\end{eqnarray*}%
and those associated with the generators $\left\{ E,\left( iE\right)
,H\right\} \ $are

\begin{eqnarray*}
g_{+}^{E} &=&\left( 
\begin{array}{cc}
\frac{1}{2}\left( \alpha \beta +\bar{\beta}\bar{\alpha}\right) \alpha -\bar{%
\beta} & \frac{1}{2}\left( \alpha \beta +\bar{\beta}\bar{\alpha}\right) \beta
\\ 
-\frac{1}{2}\left( \alpha \beta +\bar{\beta}\bar{\alpha}\right) \bar{\beta}
& \frac{1}{2}\left( \alpha \beta +\bar{\beta}\bar{\alpha}\right) \bar{\alpha}%
-\beta%
\end{array}%
\right) \\
&& \\
g_{+}^{\left( iE\right) } &=&\left( 
\begin{array}{cc}
-\frac{1}{2}\left( i\alpha \beta -i\bar{\beta}\bar{\alpha}\right) \alpha -i%
\bar{\beta} & -\frac{1}{2}\left( i\alpha \beta -i\bar{\beta}\bar{\alpha}%
\right) \beta \\ 
\frac{1}{2}\left( i\alpha \beta -i\bar{\beta}\bar{\alpha}\right) \bar{\beta}
& -\frac{1}{2}\left( i\alpha \beta -i\bar{\beta}\bar{\alpha}\right) \bar{%
\alpha}+i\beta%
\end{array}%
\right) \\
&& \\
g_{+}^{H} &=&2\left( 
\begin{array}{cc}
\left\vert \beta \right\vert ^{2}\alpha & -\left\vert \alpha \right\vert
^{2}\beta \\ 
\left\vert \alpha \right\vert ^{2}\bar{\beta} & \left\vert \beta \right\vert
^{2}\bar{\alpha}%
\end{array}%
\right)
\end{eqnarray*}

\subsubsection{Semidirect product with the $\protect\tau $-action action}

Now, we shall work with $G=SL(2,\mathbb{C})$ and its Lie algebra $\mathfrak{g%
}=\mathfrak{sl}_{2}(\mathbb{C})$ embodied in the semidirect product Lie
group $H=SL(2,\mathbb{C})$ $\ltimes \mathfrak{sl}_{2}(\mathbb{C})$ where the
vector space $\mathfrak{sl}_{2}(\mathbb{C})$ regarded as the representation
space for the $\tau $-\emph{action} in the right action structure of
semidirect product $\left( \ref{semidirect prod}\right) $. The associated
Lie algebra is the semidirect sum Lie algebra $\mathfrak{h}=\mathfrak{sl}%
_{2}(\mathbb{C})\ltimes \mathfrak{sl}_{2}(\mathbb{C})$. It inherit a
decomposition from that in $\mathfrak{sl}_{2}(\mathbb{C})$, namely, $%
\mathfrak{h}=\mathfrak{h}_{+}\oplus \mathfrak{h}_{-}$ with 
\begin{eqnarray*}
\mathfrak{h}_{+} &=&\mathfrak{g}_{+}\oplus \mathfrak{g}_{+}^{\bot }=%
\mathfrak{su}_{2}\oplus \mathfrak{b} \\
\mathfrak{h}_{-} &=&\mathfrak{g}_{-}\oplus \mathfrak{g}_{-}^{\bot }=%
\mathfrak{b}\oplus \mathfrak{su}_{2}
\end{eqnarray*}%
The bilinear form%
\begin{equation*}
\left( \left( X,X^{\prime }\right) ,\left( Y,Y^{\prime }\right) \right) _{%
\mathfrak{h}}=\mathrm{g}\left( X,Y^{\prime }\right) +\mathrm{g}\left(
Y,X^{\prime }\right)
\end{equation*}%
turns $\mathfrak{h}_{+},\mathfrak{h}_{-}$ into isotropic subspaces of $%
\mathfrak{h}$. The Lie group $H=SL(\mathbb{C})\ltimes \mathfrak{sl}_{2}(%
\mathbb{C})$ factorize also as $H=H_{+}H_{-}$ with 
\begin{equation*}
\begin{array}{ccc}
H_{+}=SU\left( 2\right) \ltimes \mathfrak{b} & , & H_{-}=B\ltimes \mathfrak{%
su}_{2}%
\end{array}%
\end{equation*}%
Then, a typical element $h\in H$ can be written as 
\begin{equation*}
h=\left( g,X\right) =\left( g_{+},\tau _{g_{-}}\Pi _{\mathfrak{g}_{+}^{\bot
}}X\right) \cdot \left( g_{-},\Pi _{\mathfrak{g}_{-}^{\bot }}X\right)
\end{equation*}

Let us determine de PL bivector on $H_{+}$ from relation 
\begin{eqnarray*}
&&\left\langle \gamma \left( X_{-}^{\prime },Y_{-}^{\prime \bot }\right)
\left( g_{+},Y_{+}^{\bot }\right) ^{-1}\otimes \gamma \left( X_{-}^{\prime
\prime },Y_{-}^{\prime \prime \bot }\right) \left( g_{+},Y_{+}^{\bot
}\right) ^{-1},\pi _{+}\left( g_{+},Z_{+}^{\bot }\right) \right\rangle \\
&:&=\left( \Pi _{-}\mathrm{Ad}_{\left( g_{+},Z_{+}^{\bot }\right)
^{-1}}^{H}\left( X_{-}^{\prime },Y_{-}^{\prime \bot }\right) ,\Pi _{+}%
\mathrm{Ad}_{\left( g_{+},Z_{+}^{\bot }\right) ^{-1}}^{H}\left(
X_{-}^{\prime \prime },Y_{-}^{\prime \prime \bot }\right) \right) _{%
\mathfrak{h}}
\end{eqnarray*}%
By writing 
\begin{eqnarray*}
X_{-} &=&x_{E}E+x_{\left( iE\right) }\left( iE\right) +x_{H}H \\
Y_{-}^{\bot } &=&y_{1}X_{1}+y_{2}X_{2}+y_{3}X_{3} \\
Z_{+}^{\bot } &=&z_{E}E+z_{\left( iE\right) }\left( iE\right) +z_{H}H
\end{eqnarray*}%
and introducing $\pi _{+}^{R}:H_{+}\longrightarrow \mathfrak{h}_{+}\wedge 
\mathfrak{h}_{+}$ such that%
\begin{eqnarray*}
&&\left\langle \gamma \left( X_{-}^{\prime },Y_{-}^{\prime \bot }\right)
\left( g_{+},Y_{+}^{\bot }\right) ^{-1}\otimes \gamma \left( X_{-}^{\prime
\prime },Y_{-}^{\prime \prime \bot }\right) \left( g_{+},Y_{+}^{\bot
}\right) ^{-1},\pi _{+}\left( g_{+},Z_{+}^{\bot }\right) \right\rangle \\
&=&\left( \left( X_{-}^{\prime },Y_{-}^{\prime \bot }\right) \otimes \left(
X_{-}^{\prime \prime },Y_{-}^{\prime \prime \bot }\right) ,\pi
_{+}^{R}\left( g_{+},Z_{+}^{\bot }\right) \right) _{\mathfrak{h}}
\end{eqnarray*}%
we obtain

\begin{eqnarray*}
&&\pi _{+}^{R}\left( g_{+}\left( \alpha ,\beta \right) ,z_{E}E+z_{\left(
iE\right) }\left( iE\right) +z_{H}H\right) \\
&=&\left\vert \beta \right\vert ^{2}\left( X_{1},0\right) \wedge \left(
0,\left( iE\right) \right) \\
&&-\left( \left( \left\vert \alpha \right\vert ^{2}-\left\vert \beta
\right\vert ^{2}\right) \mathrm{Im}\left( \beta \alpha \right) +2i\mathrm{Re}%
\left( \beta ^{3}\bar{\alpha}\right) \right) \left( X_{1},0\right) \wedge
\left( 0,H\right) \\
&&-\left( 2\mathrm{Re}\left( \alpha ^{2}\beta ^{2}\right) +\left( \left\vert
\alpha \right\vert ^{2}-\left\vert \beta \right\vert ^{2}\right) \left\vert
\beta \right\vert ^{2}\right) \left( X_{2},0\right) \wedge \left( 0,E\right)
\\
&&-2\mathrm{Im}\left( \alpha ^{2}\beta ^{2}\right) \left( X_{2},0\right)
\wedge \left( 0,\left( iE\right) \right) \\
&&+\left( \left( \left\vert \alpha \right\vert ^{2}-\left\vert \beta
\right\vert ^{2}\right) \mathrm{Re}\left( \alpha \beta \right) +2\mathrm{Re}%
\left( \beta ^{3}\bar{\alpha}\right) \right) \left( X_{2},0\right) \wedge
\left( 0,H\right) \\
&&-\mathrm{Re}\left( \alpha \beta \right) \left( X_{3},0\right) \wedge \left(
0,\left( iE\right) \right) -2\mathrm{Im}\left( \alpha ^{2}\bar{\beta}%
^{2}\right) \left( X_{3},0\right) \wedge \left( 0,H\right) \\
&&-\left( z_{E}\left( \left( 1-4\left\vert \beta \right\vert ^{2}\right) 
\mathrm{Re}\left( \alpha ^{2}\right) +\left( 1-4\left\vert \alpha \right\vert
^{2}\right) \mathrm{Re}\left( \beta ^{2}\right) \right) \right) \left(
0,E\right) \wedge \left( 0,H\right) \\
&&+z_{\left( iE\right) }\mathrm{Im}\left( \alpha ^{2}-\beta ^{2}\right) \left(
0,E\right) \wedge \left( 0,H\right) \\
&&+\left( z_{E}\left( \left( 1-4\left\vert \beta \right\vert ^{2}\right) 
\mathrm{Im}\left( \alpha ^{2}\right) +\left( 1-4\left\vert \alpha \right\vert
^{2}\right) \mathrm{Im}\left( \beta ^{2}\right) \right) \right) \left(
0,H\right) \wedge \left( 0,\left( iE\right) \right) \\
&&+z_{\left( iE\right) }\mathrm{Re}\left( \alpha ^{2}-\beta ^{2}\right) \left(
0,H\right) \wedge \left( 0,\left( iE\right) \right) \\
&&+2\left( z_{E}\mathrm{Im}\left( \alpha \bar{\beta}\right) +z_{\left(
iE\right) }\mathrm{Re}\left( \alpha \bar{\beta}\right) \right) \left( 0,\left(
iE\right) \right) \wedge \left( 0,E\right)
\end{eqnarray*}

On the side, the Poisson-Lie bivector on $H_{-}$ is defined from relation%
\begin{eqnarray*}
&&\left\langle \left( g_{-},Z_{-}^{\bot }\right) ^{-1}\gamma \left(
X_{+}^{\prime },Y_{+}^{\prime \bot }\right) \otimes \left( g_{-},Z_{-}^{\bot
}\right) ^{-1}\gamma \left( X_{+}^{\prime \prime },Y_{+}^{\prime \prime \bot
}\right) ,\pi _{-}\left( g_{-},Z_{-}^{\bot }\right) \right\rangle \\
&:&=\left( \Pi _{+}\mathrm{Ad}_{\left( g_{-},Z_{-}^{\bot }\right)
}^{H}\left( X_{+}^{\prime },Y_{+}^{\prime \bot }\right) ,\Pi _{-}\mathrm{Ad}%
_{\left( g_{-},Z_{-}^{\bot }\right) }^{H}\left( X_{+}^{\prime \prime
},Y_{+}^{\prime \prime \bot }\right) \right) _{\mathfrak{h}}
\end{eqnarray*}%
that writing%
\begin{eqnarray*}
X_{+} &=&x_{1}X_{1}+x_{2}X_{2}+x_{3}X_{3} \\
Y_{+}^{\bot } &=&b_{1}B_{1}+b_{2}B_{2}+b_{3}B_{3}
\end{eqnarray*}%
and introducing $\pi _{H}^{-}:H_{-}\longrightarrow \mathfrak{h}_{-}\wedge 
\mathfrak{h}_{-}$ such that%
\begin{eqnarray*}
&&\left\langle \left( g_{-},Z_{-}^{\bot }\right) ^{-1}\gamma \left(
X_{+}^{\prime },Y_{+}^{\prime \bot }\right) \otimes \left( g_{-},Z_{-}^{\bot
}\right) ^{-1}\gamma \left( X_{+}^{\prime \prime },Y_{+}^{\prime \prime \bot
}\right) ,\pi _{-}\left( g_{-},Z_{-}^{\bot }\right) \right\rangle \\
&=&\left\langle \left( g_{-},Z_{-}^{\bot }\right) ^{-1}\gamma \left(
X_{+}^{\prime },Y_{+}^{\prime \bot }\right) \otimes \left( g_{-},Z_{-}^{\bot
}\right) ^{-1}\gamma \left( X_{+}^{\prime \prime },Y_{+}^{\prime \prime \bot
}\right) ,\pi _{-}\left( g_{-},Z_{-}^{\bot }\right) \right\rangle
\end{eqnarray*}%
we get 
\begin{equation*}
\begin{array}{l}
\pi _{-}^{L}\left( g_{-},Z_{-}^{\bot }\right) \\ 
=\dfrac{1}{a^{2}}\left( c^{2}+b^{2}+\dfrac{1}{a^{2}}-a^{2}\right) \left(
B_{1},0\right) \wedge \left( 0,X_{2}\right) \\ 
-\dfrac{1}{a^{2}}\left( b^{2}+c^{2}+\dfrac{1}{a^{2}}-a^{2}\right) \left(
B_{2},0\right) \wedge \left( 0,X_{1}\right) \\ 
~+\dfrac{c}{a}\left( B_{1},0\right) \wedge \left( 0,X_{3}\right) -\dfrac{c}{a%
}\left( B_{3},0\right) \wedge \left( 0,X_{1}\right) \\ 
-\dfrac{b}{a}\left( B_{2},0\right) \wedge \left( 0,X_{3}\right) -\dfrac{b}{a}%
\left( B_{3},0\right) \wedge \left( 0,X_{2}\right) \\ 
~+2\dfrac{1}{a^{2}}\left( bcz_{1}-acz_{2}+2\left( c^{2}+b^{2}+\dfrac{1}{a^{2}%
}\right) z_{3}\right) \left( 0,X_{1}\right) \wedge \left( 0,X_{2}\right) \\ 
~+\dfrac{1}{a}\left( az_{1}-2bz_{3}\right) \left( 0,X_{2}\right) \wedge
\left( 0,X_{3}\right) -\dfrac{1}{a}\left( az_{2}+2cz_{3}\right) \left(
0,X_{1}\right) \wedge \left( 0,X_{3}\right)%
\end{array}%
\end{equation*}

\subsubsection{AKS integrable system on $\mathfrak{h}_{-}$}

We study a Hamiltonian system on $\mathfrak{h}_{-}$, described by the $%
\mathrm{Ad}^{H}$-invariant function $\mathsf{H}$, and for which the
hamiltonian vector field is%
\begin{equation*}
V_{\mathsf{H}}\left( X_{-}+K_{+}\right) =\Pi _{-}\mathrm{ad}_{\Pi _{+}%
\mathfrak{L}_{\mathsf{H}}\left( X_{-}+K_{+}\right) }^{\mathfrak{h}}\left(
X_{-}+K_{+}\right) =-\mathrm{ad}_{\Pi _{-}\mathfrak{L}_{\mathsf{H}}\left(
X_{-}+K_{+}\right) }^{\mathfrak{h}}\left( X_{-}+K_{+}\right)
\end{equation*}%
It is worth to recall that $\psi \left( K_{+}\right) $ must be a \emph{%
character }of $\mathfrak{h}_{-}$, which in this case means that 
\begin{equation*}
K_{+}=\left( k_{3}X_{3},k_{H}H\right)
\end{equation*}%
for arbitrary $k_{3},k_{H}\in \mathbb{R}$. Thus the Hamilton equation of
motion on $\imath _{K_{+}}\left( \mathfrak{h}_{-}\right) \subset \mathfrak{h}
$ are then%
\begin{equation*}
\dot{Z}\left( t\right) =V_{\mathsf{H}}\left( Z\left( t\right) \right)
\end{equation*}%
for the initial condition $Z\left( 0\right) =X_{-\circ }+K_{+}$.

In particular, we consider the $\mathrm{Ad}^{H}$-invariant Hamilton function%
\begin{equation*}
\mathsf{H}\left( X,X^{\prime }\right) =\mathrm{g}\left( X,X^{\prime }\right)
\end{equation*}%
where $\psi \left( \Pi _{+}\left( X,X^{\prime }\right) \right) \in \mathrm{%
char~}\mathfrak{h}_{-}$, so that $\left( X,X^{\prime }\right) =\left(
X_{-}+k_{3}X_{3},X_{-}^{\bot }+k_{H}H\right) $. Then, the Hamilton function
reduces to%
\begin{equation*}
\mathsf{H}\left( x_{E}E+x_{\left( iE\right) }\left( iE\right)
+x_{H}H+k_{3}X_{3},x_{1}X_{1}+x_{2}X_{2}+x_{3}X_{3}+k_{H}H\right)
=x_{H}k_{H}+x_{3}k_{3}
\end{equation*}%
and%
\begin{equation*}
\mathfrak{L}_{\mathsf{H}}\left( X,X^{\prime }\right) =\left(
x_{H}H+k_{3}X_{3},k_{H}H+x_{3}X_{3}\right)
\end{equation*}%
The hamiltonian vector field is 
\begin{eqnarray*}
&&V_{\mathsf{H}}\left( x_{E}E+x_{\left( iE\right) }\left( iE\right)
+x_{H}H+k_{3}X_{3},x_{1}X_{1}+x_{2}X_{2}+x_{3}X_{3}+k_{H}H\right) \\
&=&\left( -2x_{H}\left( x_{E}E+x_{\left( iE\right) }\left( iE\right) \right)
,-2\left( x_{E}x_{3}+x_{H}x_{1}\right) X_{1}+2\left( x_{\left( iE\right)
}x_{3}-x_{H}x_{2}\right) X_{2}\right)
\end{eqnarray*}%
Because $x_{H}$ and $x_{3}$ are time independent, we write them as 
\begin{equation*}
\begin{array}{ccc}
x_{H}=\dfrac{\alpha }{2} & , & x_{3}=\dfrac{\beta }{2}%
\end{array}%
\end{equation*}%
and the nontrivial Hamilton equations are%
\begin{equation}
\left\{ 
\begin{array}{l}
\dot{x}_{E}=-\alpha x_{E} \\ 
\dot{x}_{\left( iE\right) }=-\alpha x_{\left( iE\right) } \\ 
\dot{x}_{1}=-\beta x_{E}-\alpha x_{1} \\ 
\dot{x}_{2}=\beta x_{\left( iE\right) }-\alpha x_{2}%
\end{array}%
\right.  \label{ham eq sl2C}
\end{equation}

The solution $Z\left( t\right) =X_{-}\left( t\right) +K_{+}$ to this set of
Hamilton equation is%
\begin{equation*}
Z\left( t\right) =-\mathrm{Ad}_{h_{-}\left( t\right) }^{H}\left( X_{-\circ
}+K_{+}\right)
\end{equation*}%
with 
\begin{equation*}
h_{-}\left( t\right) =\left( g_{-}\left( t\right) ,tx_{30}X_{3}\right)
\end{equation*}%
being such that%
\begin{equation*}
\left( t\mathfrak{L}_{\mathsf{H}}(X_{-\circ }+K_{+})\right) =h_{+}\left(
t\right) h_{-}\left( t\right)
\end{equation*}

So, in order to solve our problem, we deal with the exponential curve%
\begin{equation*}
\mathrm{Exp}^{\cdot }t\left( x_{H0}H+k_{30}X_{3},k_{H0}H+x_{30}X_{3}\right)
=\left( e^{t\left( x_{H0}H+k_{30}X_{3}\right) },tk_{H0}H+tx_{30}X_{3}\right)
\end{equation*}%
which factorize as 
\begin{equation*}
\left( e^{t\left( x_{H0}H+k_{30}X_{3}\right) },tk_{H0}H+tx_{30}X_{3}\right)
=\left( g_{+}\left( t\right) ,tk_{H0}H\right) \cdot \left( g_{-}\left(
t\right) ,tx_{30}X_{3}\right)
\end{equation*}

The evolution of the system is then driven by the orbit of the curve $%
t\longmapsto h_{-}\left( t\right) =\left( g_{-}\left( t\right)
,tx_{30}X_{3}\right) $ where 
\begin{equation*}
g_{-}\left( t\right) =\left( 
\begin{array}{cc}
e^{tx_{H0}} & 0 \\ 
0 & e^{-tx_{H0}}%
\end{array}%
\right)
\end{equation*}%
Writing 
\begin{eqnarray*}
Z\left( t\right) &=&\left( x_{E}\left( t\right) E+x_{\left( iE\right)
}\left( t\right) \left( iE\right) +x_{H}\left( t\right) H+k_{3}X_{3}\right. ,
\\
&&\left. x_{1}\left( t\right) X_{1}+x_{2}\left( t\right) X_{2}+x_{3}\left(
t\right) X_{3}+k_{H}H\right)
\end{eqnarray*}%
and 
\begin{equation*}
Z\left( t_{0}\right) =\left( x_{E0}E+x_{\left( iE\right) 0}\left( iE\right)
+x_{H0}H+k_{3}X_{3},x_{10}X_{1}+x_{20}X_{2}+x_{30}X_{3}+k_{H}H\right)
\end{equation*}%
we have, after some computation and introducing 
\begin{equation*}
\begin{array}{ccc}
x_{H0}=\dfrac{\alpha }{2} & , & x_{30}=\dfrac{\beta }{2}%
\end{array}%
\end{equation*}%
it turns in 
\begin{eqnarray*}
Z\left( t\right) &=&-\left( x_{E0}e^{\alpha t}E+x_{\left( iE\right)
0}e^{\alpha t}\left( iE\right) +\dfrac{\alpha }{2}H+k_{3}X_{3}\right. , \\
&&\left. \left( x_{10}+\beta x_{E0}t\right) e^{\alpha t}X_{1}+\left(
x_{20}-\beta x_{\left( iE\right) 0}t\right) e^{\alpha t}X_{2}+\dfrac{\beta }{%
2}X_{3}+k_{H}H\right)
\end{eqnarray*}%
which means that%
\begin{equation*}
\left\{ 
\begin{array}{l}
x_{E}\left( t\right) =-x_{E0}e^{\alpha t} \\ 
x_{\left( iE\right) }\left( t\right) =-x_{\left( iE\right) 0}e^{\alpha t} \\ 
x_{1}\left( t\right) =-x_{10}e^{\alpha t}-\beta x_{E0}te^{\alpha t} \\ 
x_{2}\left( t\right) =-x_{20}e^{\alpha t}+\beta x_{\left( iE\right)
0}te^{\alpha t}%
\end{array}%
\right.
\end{equation*}%
that solves the equations $\left( \ref{ham eq sl2C}\right) $.

\section{Conclusions}

In this work we developed a method to obtain AKS integrable systems
overtaking the lack of an Ad-invariant bilinear form in the associated Lie
algebra. In doing so, we promoted the original double Lie algebra to the
framework of the semidirect product with the $\tau $-representation,
obtaining a double semidirect product Lie algebra which naturally admits an
Ad-invariant bilinear form. Actually, the structure is richer than this
since it amount to \emph{Manin triple}. So, the method not only brings the
problem to the very realm of AKS theory, but also to that of Lie bialgebras
and Poisson-Lie groups.

We developed these issues in extent, showing how the AKS\ theory works in
the proposed framework producing Lax pairs equations, and that by freezing
the first coordinate in the semidirect product, the $\tau $-orbits
formulation is retrieved. We built the crossed dressing actions of the
factors in the semidirect product Lie group, and showed that the dynamical
systems moves on the associated orbits. The Lie bialgebra and Poisson-Lie
structures were obtained, however we do not advanced forward the coboundary
ones leaving them a pending problem.

In the examples we explored three different situations lacking of
Ad-invariant bilinear form: a nilpotent and a solvable Lie algebras, and $%
\mathfrak{sl}_{2}\left( \mathbb{C}\right) $ where the usual Ad-invariant
bilinear form is spoiled with the aim of getting an inner product which
gives rise to a Riemannian metric on $SL\left( 2,\mathbb{C}\right) $.

In summary, it has been shown that an AKS scheme and Lie bialgebra - PL
structures can be associated to any Lie algebra without Ad-invariant
bilinear form by promoting them to the semidirect product with the $\tau $%
-representation.

\section{Acknowledgments}

The authors thank to CONICET (Argentina) for financial support.


\end{document}